\documentclass[conference]{IEEEtran}
\IEEEoverridecommandlockouts

\usepackage[T1]{fontenc}
\usepackage[utf8]{inputenc}
\usepackage{amsmath,amssymb}
\usepackage{booktabs}
\usepackage{graphicx}
\usepackage{siunitx}
\usepackage{xcolor}
\usepackage{listings}
\usepackage{algorithm}
\lstdefinestyle{pypseudo}{%
  language=Python,
  basicstyle=\scriptsize\ttfamily,
  keywordstyle=\bfseries,
  commentstyle=\itshape\color{gray},
  numbers=left, numberstyle=\tiny, numbersep=6pt,
  columns=fullflexible, keepspaces=true,
  mathescape=true, showstringspaces=false,
  xleftmargin=1.4em, aboveskip=0pt, belowskip=0pt,
}
\usepackage{pgfplots}
\pgfplotsset{compat=1.18}
\usepackage{placeins}
\usepackage[hidelinks]{hyperref}
\usepackage{cite}

\newcommand{\psweepSlopeCpp}{491}
\newcommand{\psweepSlopeRust}{295}
\newcommand{\psweepRsqCpp}{1.0000}
\newcommand{\psweepRsqRust}{0.9999}
\newcommand{\psweepCrossCpp}{\num{197}}
\newcommand{\psweepCrossRust}{\num{313}}
\newcommand{\psweepResidualCpp}{3.4}
\newcommand{\psweepResidualRust}{17.0}
\newcommand{\psweepResidualPCpp}{\num{500}}
\newcommand{\psweepResidualPRust}{\num{2000}}
\newcommand{\psweepPmin}{\num{250}}
\newcommand{\psweepPmax}{\num{8000}}
\newcommand{\psweepKdPerPointMin}{194.6}
\newcommand{\psweepKdPerPointMax}{195.8}

\newcommand{\legalWorstMaxCpp}{239.8}
\newcommand{\legalWorstMaxRust}{122.2}
\newcommand{\legalWorstKbar}{8.0}
\newcommand{\legalWorstRatioRust}{4.8}
\newcommand{\legalWorstRatioCpp}{4.1}
\newcommand{\legalWorstBudgetXRust}{1.2}
\newcommand{\legalOscMaxCpp}{193.0}
\newcommand{\legalOscMaxRust}{65.8}
\newcommand{\legalOscKbar}{3.0}

\newcommand{\regLooMaxRust}{97}
\newcommand{\regN}{16}
\newcommand{\regCorr}{0.99}

\newcommand{\realFrames}{\num{22416}}
\newcommand{\realOnMap}{501}
\newcommand{\realOffMap}{\num{21915}}
\newcommand{\realBaseAll}{\num{22183}}
\newcommand{\realShapeAll}{\num{22178}}
\newcommand{\realShapeOnMap}{263}
\newcommand{\realShapeOnMapPct}{52.5}
\newcommand{\realBaseOnly}{5}
\newcommand{\realBaseDivergent}{233}
\newcommand{\realIterationDivergent}{28}
\newcommand{\realOutputDivergent}{205}
\newcommand{\realNotShapeOnMap}{238}
\newcommand{\realNotShapeOnMapPct}{47.5}
\newcommand{\realDivPctAll}{0.1}
\newcommand{\realOnMapDivPct}{5.6}
\newcommand{\realOffMapIterMed}{0}
\newcommand{\realOffMapCppMed}{2.9}
\newcommand{\realOnMapSeqMax}{539}
\newcommand{\realIterCapPct}{73.1}
\newcommand{\realSegments}{15}
\newcommand{\realDivSegments}{3}
\newcommand{\realDivIterMin}{9}
\newcommand{\realDivIterMax}{30}
\newcommand{\realDivKbarMin}{0.1}
\newcommand{\realDivKbarMax}{1.5}
\newcommand{\realKmaxMax}{9}
\newcommand{\realPMax}{1500}
\newcommand{\realRustMax}{57.2}
\newcommand{\realCppMax}{133.3}
\newcommand{\realOverrunsCpp}{284}
\newcommand{\realOverrunsCppOnMapPct}{56.7}
\newcommand{\realCppMedOnMap}{106.5}

\newcommand{\realProdOnMap}{\num{2277}}
\newcommand{\realCaptureSpanMin}{33.5}
\newcommand{\realDupClouds}{\num{5582}}
\newcommand{\realProdOnMapSegments}{250}
\newcommand{\realProdOnMapSpanMin}{26}
\newcommand{\realProdIterCapPct}{70.0}
\newcommand{\realProdIterationDivergent}{133}
\newcommand{\realABCommonN}{501}
\newcommand{\realABCapDegPct}{73.1}
\newcommand{\realABCapProdPct}{44.1}

\newcommand{\tailsRatioMin}{0.08}
\newcommand{\tailsRatioMax}{0.77}
\newcommand{\tailsMinMedian}{7.7}
\newcommand{\tailsMaxObserved}{988.2}
\newcommand{\unifiedFixtureMaxCpp}{988.2}
\newcommand{\unifiedFixtureMaxRust}{610.5}
\newcommand{\tailsLatencySpread}{129}
\newcommand{\tailsRustFasterCount}{10}
\newcommand{\sessMedSpreadWorstPct}{6.1}
\newcommand{\sessRatioSpreadWorst}{0.03}
\newcommand{\subnormalMedianRatio}{12.7}

\newcommand{\searchZeroKdM}{12.14}
\newcommand{\searchZeroSumNbrM}{3.97}
\newcommand{\legalWorstSumNbrK}{372}
\newcommand{\twoTierKernelDrop}{10.7}
\newcommand{\allocWorstPerAlign}{\num{682248}}
\newcommand{\allocPerPtPass}{11}
\newcommand{\allocPerPtPassExact}{11.0}
\newcommand{\allocCertMin}{11.003}
\newcommand{\allocCertMax}{11.005}
\newcommand{\allocCertSpreadPct}{0.02}
\newcommand{\allocCertSparse}{9.4}

\newcommand{\evtAcfWorst}{0.74}
\newcommand{\evtFitFailures}{1}
\newcommand{\evtFitChecks}{18}
\newcommand{\evtGumbelExtraPct}{6.5}

\newcommand{\envCpu}{AMD Ryzen 9 5900HX with Radeon Graphics}
\newcommand{\envKernel}{Linux 6.8.0-136-generic}
\newcommand{\envCppCompiler}{c++ (Ubuntu 11.4.0-1ubuntu1\textasciitilde{}22.04.3) 11.4.0}
\newcommand{\envRustCompiler}{rustc 1.98.0-nightly (654079540 2026-06-12)}
\newcommand{\envCppReleaseFlags}{\texttt{-O3 -DNDEBUG}}
\newcommand{\envGovernor}{performance}
\newcommand{\envWarmup}{10}
\newcommand{\envSessions}{3}
\newcommand{\envSamplesPerSession}{1000}
\newcommand{\envPooledSamples}{3000}
\newcommand{\envCommit}{6541a276}

\newcommand{\certFixtures}{8}
\newcommand{\certScoreExact}{7}
\newcommand{\certMaxTransDelta}{\num{2e-07}}
\newcommand{\certMaxScoreDelta}{\num{6e-08}}

\newcommand{\legalKSearchMax}{6}
\newcommand{\legalKSearchSeeds}{4}

\newcommand{\raspiTotal}{8}
\newcommand{\raspiExact}{8}
\newcommand{\raspiDeviceOk}{8}

\newcommand{\coldWorstPct}{3.1}
\newcommand{\coWorstCppPct}{7.3}
\newcommand{\coWorstCppFixture}{\emph{geom-stress}}
\newcommand{\coWorstCppMode}{memory-bandwidth}
\newcommand{\coWorstRustPct}{3.0}
\newcommand{\coWorstRustFixture}{\emph{geom-stress}}
\newcommand{\coWorstRustMode}{memory-bandwidth}

\newcommand{\coPairedCells}{50}
\newcommand{\coRustLowerCells}{50}

\newcommand{\bridgeCppInflMin}{0.77}
\newcommand{\bridgeCppInflMax}{0.99}
\newcommand{\bridgeRustInflMin}{1.00}
\newcommand{\bridgeRustInflMax}{1.02}
\newcommand{\bridgeCppTaxMinMs}{9}
\newcommand{\bridgeCppTaxMaxMs}{35}
\newcommand{\bridgeGapBLegalWorst}{0.50}
\newcommand{\bridgeGapALegalWorst}{0.57}

\newcommand{\raspiTimingSamples}{100}
\newcommand{\raspiTimingCold}{20}
\newcommand{\raspiCalibDriftPct}{0.04}
\newcommand{\raspiSpreadWorstPct}{0.3}
\newcommand{\raspiColdDeltaWorstPct}{0.5}
\newcommand{\raspiSearchWarmMaxS}{9.91}
\newcommand{\raspiHostFactor}{16.8}
\newcommand{\raspiLegalWorstBudgetX}{15}
\newcommand{\raspiHostFactorMin}{12}
\newcommand{\raspiHostFactorMax}{18}

\newcommand{\traceFixtures}{18}
\newcommand{\traceUlpConvMax}{12}
\newcommand{\traceUlpWorstTex}{4.4\times 10^{12}}
\newcommand{\tracePayloadExactFixtures}{0}
\newcommand{\traceRealWorkGapFrames}{5}
\newcommand{\traceRealWorkGapNbrMed}{1}
\newcommand{\traceRealWorkGapNbrMax}{12}
\newcommand{\traceRealWorkGapRelMaxPct}{0.05}
\newcommand{\tracePayloadExactOnMap}{2}
\newcommand{\traceRealDivergentTransDeltaMax}{0.40217}

\newcommand{\parCalibDriftPct}{2.5}
\newcommand{\parRustLegalFitK}{2}
\newcommand{\parRustLegalFitMaxMs}{61.0}
\newcommand{\parRustLegalSerialMs}{122.9}
\newcommand{\parRustSpeedupFour}{3.9}

\newcommand{\subnormalShellSubCpp}{103.8}
\newcommand{\subnormalShellCtrlCpp}{111.5}
\newcommand{\subnormalShellSubRust}{7.2}
\newcommand{\subnormalShellCtrlRust}{8.0}
\newcommand{\subnormalShellSourceM}{1.99}
\newcommand{\subnormalShellControlM}{1.69}

\newcommand{\runtimePMax}{2000}
\newcommand{\runtimeLeavesObserved}{347956}
\newcommand{\runtimeLeavesMax}{418000}
\newcommand{\runtimeScratchBytes}{27488}
\newcommand{\runtimeKernelEvalMax}{\num{3968000}}
\newcommand{\runtimeKdVisitMax}{\num{25916000000}}
\newcommand{\runtimeValidationFrames}{22416}
\newcommand{\runtimeObservedKMax}{10}
\newcommand{\runtimeNeighborLimitFrames}{0}

\newcommand{\stackInputFrames}{\num{34375}}
\newcommand{\stackWorkers}{4}
\newcommand{\stackMapExtent}{500}
\newcommand{\stackMapRadius}{150}
\newcommand{\stackStarve}{25}

\newcommand{\stackCppTimeMax}{155.9}

\newcommand{\stackRustTimeMax}{25.8}
\newcommand{\stackCppExeMedMin}{17.5}
\newcommand{\stackCppExeMedMax}{17.9}
\newcommand{\stackRustExeMedMin}{7.6}
\newcommand{\stackRustExeMedMax}{8.0}
\newcommand{\stackCppSurvMin}{82.3}
\newcommand{\stackCppSurvMax}{83.2}
\newcommand{\stackRustSurvMin}{93.5}
\newcommand{\stackRustSurvMax}{96.5}
\newcommand{\stackCppReinitMin}{7}
\newcommand{\stackCppReinitMax}{11}
\newcommand{\stackRustReinitMin}{3}
\newcommand{\stackRustReinitMax}{5}
\newcommand{\stackReinitTotal}{26}
\newcommand{\stackCppCapMin}{23.7}
\newcommand{\stackCppCapMax}{27.8}
\newcommand{\stackRustCapMin}{10.3}
\newcommand{\stackRustCapMax}{10.6}
\newcommand{\stackIterMedMin}{13}
\newcommand{\stackIterMedMax}{15}

\newcommand{\stackCppOverMin}{3}
\newcommand{\stackCppOverMax}{4}
\newcommand{\stackSharedEpisodes}{2}
\newcommand{\stackCppOnlyEpisodes}{2}
\newcommand{\stackQRepeatMin}{\num{3658}}
\newcommand{\stackQRepeatMax}{\num{6714}}
\newcommand{\stackQCrossMin}{\num{9527}}
\newcommand{\stackQCrossMax}{\num{9953}}
\newcommand{\stackInnovRepeatMin}{0.91}
\newcommand{\stackInnovRepeatMax}{1.46}
\newcommand{\stackInnovCrossMin}{2.11}
\newcommand{\stackInnovCrossMax}{2.15}

\newcommand{\ndt}{NDT}
\newcommand{\maxnn}{\ensuremath{K_{\max}}}
\newcommand{\pwcet}{pWCET}
\newcommand{\sumnbr}{\ensuremath{\Sigma_{\mathrm{nbr}}}}
\newcommand{\kdnodes}{\ensuremath{\Sigma_{\mathrm{node}}^{R}}}
\newcommand{\flannevals}{\ensuremath{\Sigma_{\mathrm{FLANN}}^{C}}}

\begin{document}

\title{Conditional Work Bounds and Timing Evidence for an Industrial \ndt{} Scan Matcher:\\
Cross-Language Work-Trace Conformance}

\author{%
  \IEEEauthorblockN{Yuuki Takano}
  \IEEEauthorblockA{TIER IV, Inc.}
  \thanks{Preprint. This manuscript has not been peer reviewed.}
}

\maketitle

\begin{abstract}
Normal Distributions Transform (\ndt) scan matching has strongly input-dependent execution
time on the localization critical path. We study the align kernel of Autoware's production
matcher; map loading, preprocessing, and Robot Operating System scheduling are outside the
timed scope. Our three-layer method enforces and property-tests a conditional structural work
envelope, constructs stress witnesses with deterministic Rust counters, and measures frozen
witnesses. It separates algorithmic work from platform time and does not claim a certified
hard time bound.

A domain-informed construction attains the configured ceiling for the language-common
kernel-evaluation term; two archived candidates exercise higher Rust-specific kd-tree
traversal at lower common work. Per-input traces define \emph{work-trace conformance} over
passes, processed points, neighbor counts, and leaf identities. Leaf payloads,
implementation-specific traversal, total work, and latency remain outside that relation, so
C++ results are observations over transferred fixtures rather than a C++ worst-case search.
An auxiliary cropped-map, degraded-prior replay exposes the remaining conformance
boundary, and a closed-loop replay with recovery shows both engines re-initializing at the
same route sections with disjoint coverage bands.

In a unified three-boot synthetic campaign, the largest observed values are
\SI{\unifiedFixtureMaxCpp}{ms} for C++ and \SI{\unifiedFixtureMaxRust}{ms} for Rust; Rust
has the smaller maximum on all \tailsRustFasterCount{} paired fixtures. On
neighbor-returning fixtures, C++ performs approximately \allocPerPtPass{} heap allocations
per point per derivative pass, whereas the Rust align path allocates none within its reserved
envelope. Rust narrows the safe-core verification surface but is not required for bounded
execution; its unsafe code, foreign-function interface, and dependencies remain trusted.
\end{abstract}


\section{Introduction}
\label{sec:intro}

Autoware~\cite{autoware} localizes a vehicle by aligning each light-detection-and-ranging (LiDAR) scan to a Gaussian
voxel map with the Normal Distributions Transform (\ndt)~\cite{biber2003ndt}. At 10\,Hz,
each alignment has a \SI{100}{ms} period before a late pose forces downstream estimation toward
dead reckoning. Yet input geometry changes Newton iterations, neighbor multiplicity, and
kd-tree traversal. Across the synthetic stress inputs, the lowest engine--fixture median is
\SI{\tailsMinMedian}{ms}, whereas the largest observed latency is
\SI{\tailsMaxObserved}{ms} on the same core, a \tailsLatencySpread-fold span. Nominal
replay therefore cannot answer whether the deadline has defensible margin.

Static and parametric worst-case-execution-time (WCET) analysis can express
input-dependent bounds, but this kernel
still requires a realizable geometry that jointly exercises its dominant terms. Exhaustive
enumeration is infeasible, while wall-clock-guided search is noisy and machine-specific. We
instead separate a structural work envelope from platform timing and search deterministic
work counters for stress witnesses.

Rust is useful here not because C++ cannot be bounded, but because ownership, lifetimes, and
aliasing rules discharge memory-safety and data-race-freedom obligations in the safe
core~\cite{jung2018rustbelt}. This narrows project-specific validation to equivalence,
numerics, work bounds, and timing. The reduction excludes the FFI, unsafe code, and trusted
dependencies. The port also builds as \texttt{no\_std} for real-time environments without
the standard library. Its radius query caps neighbors at \maxnn; the
\emph{bounded-neighbor domain} $D$ contains inputs whose uncapped queries already satisfy
that limit. Per-input work traces compare a defined work shape on evaluated inputs; leaf
payloads, traversal, memory operations, and latency are assessed separately.

Our objective is to make the \ndt{} align kernel amenable to real-time analysis through
explicit work limits and timing evidence. The Rust port is both an implementation and an
analysis instrument; the production C++ engine is its behavioral and timing baseline. We do
not search for or establish the C++ implementation's worst traversal shape.

Our contributions are:
\begin{enumerate}
  \item A three-layer analysis combining a property-tested structural work bound,
    deterministic stress-witness generation using Rust counters, and controlled timing of
    frozen witnesses.
  \item A per-input cross-language work-trace conformance check. The base-agreement set
    $E_{\mathrm{base}}$ checks iteration equality and pose and score tolerances;
    $E_{\mathrm{shape}}$ additionally checks the recorded
    pass structure, point and neighbor counts, and canonical leaf identities. A separate
    payload digest checks leaf means and inverse covariances. Neither establishes runtime
    membership in $D$ or full semantic equivalence.
  \item A domain-informed construction that reaches the configured maximum of
    \runtimeKernelEvalMax{} common kernel evaluations, together with two archived candidates that
    exercise high Rust kd-node counts at lower common-kernel work under the same timing
    protocol.
  \item An industrial case study in which Rust removes the C++ per-point allocation hazard
    within reserved capacities and has a smaller observed maximum on every paired fixture in
    the unified three-boot campaign. Contract-geometry and
    shipped-configuration witnesses separate production input constraints from
    configuration effects, while open- and closed-loop replays of a single drive expose
    the residual conformance boundary and engine-level recovery behavior.
\end{enumerate}

Section~\ref{sec:problem} defines the work model;
Secs.~\ref{sec:methodology}--\ref{sec:implementation} present the method and port.
Section~\ref{sec:evaluation} evaluates the evidence, followed by limitations, related work,
and conclusions. The appendix contains secondary diagnostics.

\section{Problem: WCET of an Iterative Scan Matcher}
\label{sec:problem}

\subsection{The NDT align kernel}
\ndt{} registration~\cite{biber2003ndt,magnusson2009thesis} represents map voxels by
regularized Gaussian means and covariances. Each Newton step transforms all $P$ scan points,
radius-searches a kd-tree of nearby voxel centroids, accumulates score, gradient, and a
$6{\times}6$ Hessian over point--leaf pairs, solves the Newton system, and updates the
six-degree-of-freedom pose. Execution stops at update threshold $\epsilon$ or the
30-iteration cap.

We study the serial Autoware multigrid engine, based on the Point Cloud Library (PCL) and derived from
\texttt{ndt\_omp}~\cite{autoware,ndtomp}. Dynamic map tiles retain separate voxel grids,
while one kd-tree spans all leaf centroids; overlapping tiles can therefore multiply
neighbors. The \emph{engine} is the align kernel and map structures. The ROS \emph{node}
provides scans, initial guesses, and tiles and consumes poses; it is outside the timing
kernel.

\subsection{Execution-time decomposition and notation}
\label{sec:decomposition}
For one align invocation, let $T_{\mathrm{model}}$ denote the modeled kernel time, $P$ the
number of source points, and $p\in\{1,\ldots,P\}$ a source-point index. Let
$N_{\mathrm{iter}}$ be the number of completed Newton updates and $N_{\mathrm{pass}}$ the
number of derivative-evaluation passes. For point $p$, $T_{\mathrm{search}}(p)$ bounds one
radius search and $K(p)$ bounds the number of returned leaves. Finally,
$T_{\mathrm{kernel}}$ bounds one point--leaf score, gradient, and Hessian evaluation, and
$T_{\mathrm{solve}}$ bounds one Newton solve and pose update. We use the following
parametric decomposition:
\begin{equation}
 T_{\mathrm{model}}\le N_{\mathrm{pass}}\sum_{p=1}^{P}
 \bigl(T_{\mathrm{search}}(p)+K(p)T_{\mathrm{kernel}}\bigr)
 +N_{\mathrm{iter}}T_{\mathrm{solve}}.
 \label{eq:decomp}
\end{equation}
The pass index is suppressed by taking $T_{\mathrm{search}}(p)$ and $K(p)$ as
maxima over the passes of the invocation. These $T_*$ terms are symbolic unit-cost bounds,
not coefficients inferred from the timing measurements. We write
$K=\max_{1\le p\le P}K(p)$ for the maximum neighbor multiplicity over the complete
invocation.

The shipped configuration disables line-search derivative refinement, so one initial pass
plus one pass per Newton step gives $N_{\mathrm{pass}}=N_{\mathrm{iter}}+1\le31$.
The bounded Rust engine enforces $P\le P_{\max}$,
$N_{\mathrm{leaves}}\le L_{\max}$, and $K(p)\le\maxnn=64$. Here, $P_{\max}$ is
the configured source-point limit, $N_{\mathrm{leaves}}$ is the number of active map
leaves indexed by the kd-tree, $L_{\max}$ is its configured leaf limit, and \maxnn{} is
the per-query neighbor cap. A kd traversal visits at most $N_{\mathrm{leaves}}$ nodes per
query. The principal evaluation configuration uses $P_{\max}=\runtimePMax$ and
$L_{\max}=\runtimeLeavesMax$: production replay observes at most \realPMax{} scan points, while
the engine stress fixtures use 2000. Section~\ref{sec:eval-psweep} separately varies the
configured $P_{\max}$. Table~\ref{tab:preconditions} records the principal limits and
violation behavior.

We count total point--leaf evaluations as \sumnbr, total Rust kd nodes as \kdnodes, and
$\bar K=\sumnbr/(P N_{\mathrm{pass}})$. For C++, we aggregate the Fast Library for Approximate Nearest Neighbors (FLANN)
traversal events as
$\flannevals=\Sigma_{\mathrm{dist}}^{C}+\Sigma_{\mathrm{plane}}^{C}$, the sum of
distance and split-plane evaluations. Thus \sumnbr{} is included in the work-trace
conformance check, whereas \kdnodes{} and \flannevals{} count different events and their
fitted costs are engine-specific. A diagnostic timing model adds fixed per-point and
$6{\times}6$ solve terms. Equation~\eqref{eq:decomp} is therefore a parametric envelope for
the modeled kernel work, not a bound on total elapsed time: fixed setup, point
transformation, memory-system interference, and operating-system delay require separate
platform-specific bounds.

\subsection{Why work bounds need timing evidence}
Individual maxima need not be jointly reachable, so constructed inputs are witnesses, not
upper bounds. Cache, allocator, and floating-point costs are absent from the work model.
In particular, C++ performs about \allocPerPtPass{} heap allocations per point per derivative pass on
neighbor-returning fixtures. This is a known input-proportional latency path whose cost is
not bounded by the work model;
removing it eliminates one such path but does not by itself establish a finite latency bound
for the kernel or the surrounding Linux system.

\section{Methodology}
\label{sec:methodology}

Three analysis layers establish a conditional work bound, search for Rust-counter stress
inputs, and measure frozen witnesses. Per-input cross-language traces check recorded work-trace
conformance before timing, and a cross-instruction-set-architecture (cross-ISA) replay checks transfer to the
embedded evaluation target.

\subsection{Analysis layer 1: Structural bound and property tests}
\paragraph{Conditional work-envelope proposition.}
Let $P_{\max}$ and $L_{\max}$ be positive limits supplied at engine construction, and
let $I_{\max}\in\{0,\ldots,30\}$ be the configured iteration limit. For one
alignment, $P$ is the source-point count, $N_{\mathrm{leaves}}$ is the active map-leaf
count, and $N_{\mathrm{iter}}$ is the number of completed Newton updates. Let
$N_{\mathrm{pass}}$ count derivative-evaluation passes, including the initial evaluation,
and let $K(p,i)$ be the number of leaves evaluated for source point $p$ in pass $i$.
Finally, $\sumnbr$ is the total number of point--leaf kernel evaluations and $\kdnodes$
is the total number of Rust kd-tree nodes visited. Suppose $P\le P_{\max}$,
$N_{\mathrm{leaves}}\le L_{\max}$, $N_{\mathrm{iter}}\le I_{\max}$, and
$K(p,i)\le\maxnn=64$ for every
$p\in\{1,\ldots,P\}$ and $i\in\{1,\ldots,N_{\mathrm{pass}}\}$. Assume line-search
refinement is disabled, the balanced kd-tree fits
the 64-entry query stack, all scalar and point inputs are finite, and caller-owned scratch
has been reserved for $(P_{\max},I_{\max})$. Checked construction also requires the
integer products below to fit in their counter types. Then the serial Rust align kernel
satisfies
\begin{equation}
 \begin{aligned}
 N_{\mathrm{pass}}&\le I_{\max}+1,\qquad
 \sumnbr\le(I_{\max}+1)P_{\max}\maxnn,\\
 \kdnodes&\le(I_{\max}+1)P_{\max}L_{\max}.
 \end{aligned}
 \label{eq:rust-envelope}
\end{equation}
For the principal configuration $P_{\max}=\runtimePMax$,
$L_{\max}=\runtimeLeavesMax$, and $I_{\max}=30$, these bounds are
\runtimeKernelEvalMax{} kernel evaluations and \runtimeKdVisitMax{} kd-node visits. The project-owned serial kernel performs no heap allocation,
lock acquisition, or recursive call after entry. Capacity, arithmetic, stack, iteration,
map-size, and non-finite-input violations return typed errors. A query that finds a 65th
in-radius leaf retains the first 64, records a diagnostic, and forces the result to
non-converged so the node can issue a warning and reject convergence. Fixed-size linear
algebra and other dependencies remain trusted; this proposition is not a formal proof of
their internals or of elapsed time.

The instrumented counters \(\mathrm{passes}\) and \(\mathrm{points}\) denote the
derivative-pass count and total processed point--pass pairs. Randomized property tests check
$\mathrm{passes}\le N_{\mathrm{iter}}+1$,
$\mathrm{points}=P\,\mathrm{passes}$,
$\sumnbr\le\mathrm{points}\,\maxnn$, and
$\kdnodes\le\mathrm{points}\,N_{\mathrm{leaves}}$. A counting allocator verifies zero
allocations from the first frame inside the declared envelope. Linting excludes explicit
panic paths in project code, and differential tests compare the fixed-stack query against
a recursive test oracle.

\subsection{Analysis layer 2: Stress-witness generation with Rust counters}
\label{sec:meth-search}
Compile-time instrumentation in the Rust port counts passes, points, \sumnbr, and
\kdnodes; disabled counters compile away. A seeded hill-climb mutates tile overlap and
extent, centroid geometry, source composition, initial guess, and convergence threshold to
maximize $(N_{\mathrm{iter}},\sumnbr,\kdnodes)$ lexicographically. This
\emph{engine-configuration stress} search jointly varies input geometry and configuration;
it is not an input-only search under shipped parameters. $P$ remains a caller contract,
not a search dimension. A separate \emph{shipped-configuration stress} search fixes
$\epsilon=0.01$ and enforces the production preprocessing input contracts. The counter
fitness is independent of wall-clock timing and deterministic for the fixed Rust
implementation. Because lexicographic order need not match time order, two additional
high-traversal candidates at lower common-kernel work are retained as stress inputs and
timed under the same protocol. We do not compare
search strategies because the exploratory arms did not share initial distributions or provide
sufficient replication.

Only \sumnbr{} is a common work axis across the two engines. The \kdnodes{} objective is
specific to the Rust kd-tree; the search does not optimize the C++ FLANN distance or
plane-evaluation counters. Replaying the resulting fixtures on C++ provides a port baseline
and timing observations over the transferred set, not evidence of a C++ traversal extremum.

\subsection{Cross-language transfer: Per-input work-trace conformance}
\label{sec:meth-transfer}
The shipped C++ sources remain byte-identical. A local analysis build and the Rust port
instead emit matched per-pass traces; the same fixture bytes then run on the unmodified C++
binary for timing. Floating-point streams need not be bit-identical because FLANN returns
distance-sorted neighbors while Rust uses kd-visit order.

Three definitions limit the comparison. The \textbf{bounded-neighbor domain} $D$ contains
inputs for which every radius query generated by the capped Rust execution would return at
most \maxnn{} leaves if evaluated without the cap at the same query coordinate. The C++
analysis build performs uncapped searches and records each complete returned list before
kernel evaluation.

Subscripts $C$ and $R$ denote the C++ analysis build and the Rust engine. For an evaluated
input fixture $x$, let $N_{\mathrm{iter}}^e$, $t_e$, $q_e$, $\pi_e$, and $\lambda_e$
denote, respectively, engine $e\in\{C,R\}$'s completed Newton updates, output
translation, normalized output quaternion, transformation probability, and nearest-voxel
transformation likelihood. Suppressing the dependence on $x$, define
\[
 \Delta_t=\lVert t_C-t_R\rVert_2,\qquad
 \Delta_\theta=2\arccos |\langle q_C,q_R\rangle|,
\]
with $\Delta_{\mathrm{tp}}=|\pi_C-\pi_R|$ and
$\Delta_{\mathrm{nvtl}}=|\lambda_C-\lambda_R|$. Let
$J=\{t,\theta,\mathrm{tp},\mathrm{nvtl}\}$ index the four comparison metrics, and let
$\tau_j$ denote the threshold for metric $j\in J$. The \textbf{base-agreement set} is
\[
 E_{\mathrm{base}}=\{x:N_{\mathrm{iter}}^C=N_{\mathrm{iter}}^R\
 \ \text{and}\ \Delta_j\le\tau_j\ \text{for every}\ j\in J\}.
\]
The thresholds are
$\tau_t=10^{-4}\,\mathrm{m}$, $\tau_\theta=10^{-5}\,\mathrm{rad}$, and
$\tau_{\mathrm{tp}}=\tau_{\mathrm{nvtl}}=10^{-4}$.
All quantities must also be finite. These fixed numerical-agreement thresholds are much
smaller than operational localization tolerances; they are comparison criteria, not
production acceptance limits.

Each map leaf receives a canonical identifier $(g,v)$, where $g$ is the ordinal of its
grid in the deterministically sorted map and $v$ is its signed voxel key. Before hashing,
each point's returned leaves are sorted by this identifier. A domain-separated SHA-256
digest preserves point and pass order while recording the sorted identifier list. Let $i$
index derivative-evaluation passes, $P_i$ be the source-point count in pass $i$,
$S_{\mathrm{nbr},i}$ its point--leaf evaluation count, and
$\mathbf H_{\mathrm{ID},i}$ the point-ordered sequence of canonical-ID digests. The work
trace for one engine is
\[
 W_{\mathrm{shape}}=
 \left(N_{\mathrm{pass}},
 \left(P_i,S_{\mathrm{nbr},i},\mathbf H_{\mathrm{ID},i}\right){}_
 {i=1}^{N_{\mathrm{pass}}}\right).
\]
The \textbf{trace-conformant set} is
\[
 E_{\mathrm{shape}}=
 \{x\in E_{\mathrm{base}}:
 W_{\mathrm{shape}}^C(x)=W_{\mathrm{shape}}^R(x)\}.
\]
A second SHA-256 payload digest includes each leaf ID, mean, and regularized inverse
covariance. It is reported independently and is not part of $E_{\mathrm{shape}}$.
This distinction prevents shape equality from being interpreted as equality of all
kernel operands.

Because the C++ search is uncapped, membership in $E_{\mathrm{shape}}$ shows that every
corresponding C++ query returned at most \maxnn{} leaves. It does not establish membership
in $D$, however, because floating-point differences can give the engines slightly
different transformed points and query coordinates. A C++ query can return 64 leaves while
the Rust coordinate admits a 65th; the capped Rust trace still contains only its first 64.
The Rust engine separately reports that 65th leaf as a runtime diagnostic; the diagnostic
is not part of $E_{\mathrm{shape}}$.

Work-trace conformance is a dynamic check on each evaluated input. It does not imply full
semantic equivalence, membership in $D$, or equal total work. SHA-256 collision risk is
computational rather than a formal impossibility, and the selected payload still does not
record every execution state. C++ FLANN traversal, result containers, and allocation
behavior, together with Rust kd-tree traversal and flat-buffer operations, are outside
$W_{\mathrm{shape}}$; each engine therefore
has a separate work-model instance. A conformant transferred fixture is a shared timing
witness, not proof of equal stress.

\subsection{Analysis layer 3: Controlled measurement of frozen witnesses}
Frozen little-endian fixtures feed both engines through one replay harness in documented
host configurations. An \texttt{LD\_PRELOAD} interposer counts C++ allocations without
source changes. Regression against engine-specific counters is diagnostic only. Tail-model
fitting is also secondary to the observed timing results. Per-session
peaks-over-threshold/generalized-Pareto-distribution (POT/GPD) fits use maximum likelihood
at a fixed 95\% threshold, but a mandatory gate requires independence and in-sample
adequacy before any extrapolation may be reported.

\subsection{Cross-ISA transfer: Embedded-target validation}
\label{sec:meth-crossisa}
Pure-Rust \texttt{libm} removes dependence on the platform C library. In an auxiliary
cross-ISA experiment, we replay the frozen fixtures on a Raspberry Pi~4 bare-metal
\texttt{no\_std} kernel and compare each counter
tuple with its host value rather than assuming ISA independence. Section~\ref{sec:eval-raspi4}
reports the results, and Appendix~\ref{app:crossisa-controls} documents the consistency controls.

\section{Implementation}
\label{sec:implementation}

\subsection{Engines and real-time hardening}
The baseline is serial PCL/OpenMP multigrid \ndt{}~\cite{autoware,autowarecore,ndtomp}. The Rust port
is connected to the node through a bounded C application binary interface (ABI) and also
builds as \texttt{no\_std}. The constructor receives explicit work limits, and each align
call receives caller-owned scratch. This bounded ABI is an internal integration boundary.
The Rust-backed node retains the evaluated ROS-facing behavior, which does not imply
closed-loop equivalence. Differential tests cover voxel construction, align,
engine, and node paths. The primary unified campaign and $P$-sweep use
autoware\_core commit \texttt{\envCommit}.
The Rust engine source, documentation, frozen-fixture generator, and short Rust-only
evaluation script are publicly available at commit \texttt{12f9e9a} in the
\texttt{realtime\_core} repository~\cite{tier4realtimecore}.

The following are explicit constraints, not automatic Rust or \texttt{no\_std} properties:
\begin{itemize}
 \item engine construction requires positive, representable $P_{\max}$ and $L_{\max}$,
 and $I_{\max}\in\{0,\ldots,30\}$. The principal configuration sets them to \runtimePMax,
 \runtimeLeavesMax, and 30, respectively; the capacity sweep varies $P_{\max}$. For the
 principal serial build, caller-owned align scratch reserves \runtimeScratchBytes{} bytes
 for $P_{\max}$, \maxnn, and $I_{\max}+1$ result and convergence-history slots. An undersized workspace or oversized scan returns an error without growth;
 \item radius search uses a fixed 64-entry iterative stack, preserves near/far visit order,
 and is differential-tested against a recursive oracle. Finding a 65th neighbor sets a
 diagnostic while retaining the first 64;
 \item linting excludes panicking operations, align takes no locks, and disabled counters
 compile away. Checked integer products reject an invalid envelope at construction. Map
 assembly remains a control-plane operation: it may allocate, but it checks $L_{\max}$ and
 publishes only a complete map.
\end{itemize}

The active-leaf limit was sized by replaying a complete map epoch. The observed map had
\runtimeLeavesObserved{} leaves; adding 20\% and rounding upward gives
$L_{\max}=\runtimeLeavesMax$. Real replay observed $P=1500$, while the configured
$P_{\max}=\runtimePMax$ also admits the engine stress fixtures.

\subsection{Neighbor cap and engine-configuration fixtures}
The \maxnn=64 cap makes the kernel term finite. Its production-input justification assumes one
leaf per occupied voxel per tile, disjoint tiles, radius equal to voxel size, and centroids
inside their voxels. A search ball then intersects at most $3^3=27$ cells, leaving a
$2.4\times$ margin. Boundary inclusion and floating-point radius comparison remain
assumptions. A validation replay of all \runtimeValidationFrames{} captured frames
observed $K\le\runtimeObservedKMax$ and triggered the neighbor-limit diagnostic on
\runtimeNeighborLimitFrames{} frames. Overlapping tiles can exceed 64 outside these
conditions. Rust detects the 65th neighbor, returns the pose with a non-fatal
diagnostic, and forces the convergence verdict to false. The node emits a warning and treats
the frame as non-converged, allowing the existing downstream fallback policy to apply.

\label{sec:impl-fixtures}
Four hand-built engine fixtures isolate hazards: \emph{dense-nbrs} reaches 64 leaves
with eight overlapping corner lattices; \emph{max-iters} reaches the iteration cap;
\emph{cache-hostile} spreads shuffled queries across a wide map; and \emph{subnormal}
places kernel exponentials in the f64 subnormal range. The search generator exposes these
stressor dimensions. Its domain-informed seed saturates the common kernel count, and
the subsequent deterministic search explores the Rust traversal count. The per-tile $K=8$
value belongs only to the corner-lattice witness, not the analytic ceiling of 27.

\subsection{Production-input stress classes}
\label{sec:impl-legal}
Production preprocessing downsamples and crops scans, rejects duplicates, limits $P$ to
1500, and assumes disjoint voxel-aligned map tiles. The \emph{contract-geometry stress}
fixture, displayed as \emph{geom-stress}, satisfies these input conditions with $K=8$ but
uses non-shipped $\epsilon=10^{-10}$ to reach 30 iterations. At the shipped
$\epsilon=0.01$, the same fixture converges before the iteration cap.
It is therefore not a shipped-configuration execution. An independent input-contract
verifier checks each candidate.

A separate shipped-configuration search produces \emph{shipped-osc}, which satisfies the
same input contracts and reaches 30 iterations at $\epsilon=0.01$ but has
$\bar K=\legalOscKbar$. Smooth high-$K$ geometry converges quickly, while oscillatory rough
geometry has fewer neighbors, so no searched shipped-configuration input combines both
axes. \emph{geom-stress} is the slower measured run, but only \emph{shipped-osc}
establishes reachability of the iteration cap under shipped parameters. Neither is claimed
to be the worst input of its class.

\section{Evaluation}
\label{sec:evaluation}

\subsection{Setup}
\label{sec:eval-setup}
We use three named measurement configurations. \emph{Isolated} denotes controlled engine
characterization on a pinned core. \emph{Replay} denotes replay under the Linux Completely
Fair Scheduler (CFS), and \emph{Bare-metal} denotes evaluation on the AArch64
\texttt{no\_std} target. The unified serial campaign and the $P$-sweep use Isolated;
the bridge, host-scaling, real-data, and AArch64 subsections report auxiliary,
artifact-pinned experiments. They are not pooled with, or used to adjust, the primary
bounded-build timing results. The host uses \envCpu{} and \envKernel{}. The benchmark core is
isolated with \texttt{isolcpus} and \texttt{nohz\_full}; interrupt-request (IRQ)
affinities are moved, with
three immovable IRQs remaining. Its simultaneous-multithreading (SMT) sibling is offline, and the \envGovernor{}
governor pins its frequency at \SI{3.2}{GHz}. We compile C++ with \envCppCompiler{} using
\texttt{CMAKE\_BUILD\_TYPE=Release} and \envCppReleaseFlags. Rust uses
\envRustCompiler{} with Cargo release defaults \texttt{opt-level=3},
\texttt{lto=false}, and \texttt{codegen-units=16}, plus the workspace override
\texttt{overflow-checks=true}. No target CPU is explicitly selected, and the manifests
record no experiment-specific \texttt{RUSTFLAGS}. Timing binaries use
\texttt{NDT\_BUILD\_TRACED=OFF}; traced analysis binaries are never timed.

The campaign comprises \envSessions{} sessions from distinct boots, with randomized
fixture and engine order. For each fixture and engine, every session records
\envSamplesPerSession{} measured alignments after \envWarmup{} warmups, yielding
\envPooledSamples{} pooled samples per fixture and engine. We pool these samples across
sessions for the warm series, and Sec.~\ref{sec:eval-tails} summarizes the matched-size
cold and co-runner series, in which a co-runner is a concurrent interference workload.
A \texttt{steady\_clock} measures only the align loop, after the map
and kd-tree have been built once per engine and fixture. A fixed-work calibration guard
rejects a cell whose sustained speed shifts by more than 10\%. Rejected cells are
remeasured, and the pooled data contain accepted cells only. The primary campaign engines
were measured at autoware\_core commit \texttt{\envCommit}. The released data files embed per-run
manifests containing the environment, compiler flags, commits, and fixture and binary hashes.
We use p50 and p99 for the empirical 50th and 99th percentiles, respectively.

\subsection{Stress-witness outcomes}
\label{sec:eval-search}
The domain-informed seed uses the corner-lattice construction represented by
\emph{dense-nbrs}, but raises the source count from 1,500 to the engine limit of 2,000.
It saturates the analytic maximum of the Rust envelope for the common kernel-evaluation
term: $\sumnbr = P \cdot \maxnn \cdot (N_{\mathrm{iter}}{+}1)
= 2000 \cdot 64 \cdot 31 = \runtimeKernelEvalMax{}$. This construction is therefore
kernel-count-optimal under the configured $P_{\max}$, \maxnn, and $I_{\max}$.
The deterministic search then explores the Rust-specific traversal term \kdnodes{};
\emph{search-00} records $\searchZeroKdM\times10^6$ kd-node visits. In the unified
campaign it has the largest C++ observed maximum and p50. Rust p50 is largest on
\emph{pareto-01}, whereas the Rust observed maximum is largest on \emph{pareto-02}.
These orderings apply only to the evaluated fixture set. Table~\ref{tab:counters} lists
the deterministic counters recomputed at the fixed timing commit.

\begin{table}[t]
\centering
\caption{Deterministic Rust work counters of the frozen fixtures: search-00 reaches the analytic kernel-evaluation ceiling exactly ($P=2000$ for the search outputs).}
\label{tab:counters}
\footnotesize
\setlength{\tabcolsep}{4pt}
\begin{tabular}{lrrrrr}
\toprule
fixture & $N_{\mathrm{iter}}$ & \sumnbr{} & \kdnodes{} & $N_{\mathrm{leaves}}$ & tiles \\
\midrule
\emph{search-00} & 30 & \num{3968000} & \num{12140437} & \num{41472} & 8 \\
\emph{search-01} & 30 & \num{2321941} & \num{11070038} & \num{51200} & 8 \\
\emph{pareto-01} & 30 & \num{3809774} & \num{12139294} & \num{41472} & 8 \\
\emph{pareto-02} & 30 & \num{3625153} & \num{12156351} & \num{41472} & 8 \\
\emph{dense-nbrs} & 30 & \num{2976000} & \num{7778334} & \num{18432} & 8 \\
\emph{max-iters} & 30 & \num{101760} & \num{1319894} & \num{3200} & 1 \\
\emph{cache-hostile} & 3 & \num{23776} & \num{540048} & \num{28800} & 1 \\
\emph{subnormal} & 30 & \num{21198} & \num{493644} & \num{1200} & 1 \\
\emph{geom-stress} & 30 & \num{372000} & \num{4474819} & \num{92256} & 1 \\
\emph{shipped-osc} & 30 & \num{139772} & \num{3221237} & \num{26908} & 1 \\
\bottomrule
\end{tabular}
\par\smallskip\footnotesize $\sumnbr = \num{3968000} = 2000 \cdot 64 \cdot 31$ on \emph{search-00}: the Rust envelope's analytic maximum for the common kernel-evaluation term, reached exactly. \emph{search-00} jointly stresses engine input and configuration; \emph{geom-stress} uses production-contract geometry with non-shipped $\epsilon$; \emph{shipped-osc} also fixes shipped $\epsilon$.
\end{table}

\paragraph{Archived search candidates.} The fixture names \emph{pareto-01} and
\emph{pareto-02} record their origin in the exploratory search archive; they do not claim
that both remain Pareto-optimal under the final implementation. With the recomputed
counters, \emph{pareto-01} is dominated by \emph{search-00} on the recorded
$(\sumnbr,\kdnodes)$ pair, while \emph{pareto-02} trades fewer kernel evaluations for
more Rust kd-node visits. All three were retained before timing and use the unified protocol
of Sec.~\ref{sec:eval-tails}. Their different latency orderings show that neither the
recorded counter order nor an observed maximum establishes a general latency order. The
counter vector omits empty-query frequency, fixed per-point overhead, and solver behavior.

\subsection{Engine-tier frame times: C++ vs.\ Rust}
\label{sec:eval-tails}
Every fixture satisfies the work-trace criterion, although its payload digest differs.
The base diagnostics also
show identical \texttt{iteration\_num} per engine pair, final poses agreeing to at most
\certMaxTransDelta\,\si{m} of translation, and both convergence scores bit-identical on
\certScoreExact{} of \certFixtures{} fixtures.
The largest score delta is \certMaxScoreDelta{} on \emph{subnormal}, or one \texttt{f32} unit in the last place (ULP).
The timing comparison therefore uses trace-conformant inputs.
Table~\ref{tab:tails} reports the unified ten-fixture campaign, including the two archived
search candidates. Rust has a smaller observed maximum on all \tailsRustFasterCount{}
fixtures, with per-fixture ratios spanning \tailsRatioMin--\tailsRatioMax. The
engine-specific fixture maxima are \SI{\unifiedFixtureMaxCpp}{ms} for C++ and
\SI{\unifiedFixtureMaxRust}{ms} for Rust. They occur on \emph{search-00} and
\emph{pareto-02}, respectively.

\begin{table}[t]
\centering
\caption{Unified-campaign frame times in \si{ms}: Rust has the smaller observed maximum on every fixture. Serial, isolated pinned core; \envPooledSamples{} pooled samples per fixture and engine. Each fixture satisfies work-trace conformance.}
\label{tab:tails}
\footnotesize
\setlength{\tabcolsep}{4pt}
\begin{tabular}{lrrrrr}
\toprule
fixture & \multicolumn{2}{c}{C++ (p50 / max)} & \multicolumn{2}{c}{Rust (p50 / max)} & ratio \\
\midrule
\emph{search-00} & 978.3 & 988.2 & 582.5 & 588.1 & \textbf{0.60} \\
\emph{search-01} & 691.4 & 698.2 & 515.3 & 535.2 & \textbf{0.77} \\
\emph{pareto-01} & 910.8 & 914.8 & 593.2 & 598.3 & \textbf{0.65} \\
\emph{pareto-02} & 893.0 & 909.0 & 585.8 & 610.5 & \textbf{0.67} \\
\emph{dense-nbrs} & 664.9 & 668.0 & 407.3 & 424.5 & \textbf{0.64} \\
\emph{max-iters} & 146.3 & 147.0 & 31.6 & 31.9 & \textbf{0.22} \\
\emph{cache-hostile} & 32.0 & 32.4 & 7.7 & 8.0 & \textbf{0.25} \\
\emph{subnormal} & 102.2 & 103.1 & 8.0 & 8.2 & \textbf{0.08} \\
\emph{geom-stress} & 235.3 & 239.8 & 121.3 & 122.2 & \textbf{0.51} \\
\emph{shipped-osc} & 192.0 & 193.0 & 64.7 & 65.8 & \textbf{0.34} \\
\bottomrule
\end{tabular}
\par\smallskip\footnotesize ratio = Rust max / C++ max; values below one favor Rust. Isolated: 3 sessions, 2026-07-20/2026-07-21, isolated pinned core (\texttt{taskset -c 2}), 3000 measured aligns per fixture and engine.
\end{table}

The archived search-candidate rows use the same $3{\times}1000$ protocol as every other fixture. Their Rust
p50 and maximum orderings differ as described above, so neither the observed maximum alone
nor the recorded counter order establishes a general latency ordering.

A controlled check does not support subnormal arithmetic as the cause of the
\emph{subnormal} fixture's p50 ratio of $\subnormalMedianRatio\times$. Moving the source
shell from \SI{\subnormalShellSourceM}{m} to \SI{\subnormalShellControlM}{m} takes every
kernel $\exp$ result out of the subnormal band without reducing p50 for either engine: C++
changes from \SI{\subnormalShellSubCpp}{ms} to \SI{\subnormalShellCtrlCpp}{ms}, and Rust
from \subnormalShellSubRust{} to \subnormalShellCtrlRust. The residual is instead
consistent with per-point overhead in the C++ PCL/FLANN query path, including allocation and
call indirection that the allocation-free Rust query avoids. With hardware performance
counters unavailable in the measurement container, this remains mechanistic interpretation
rather than a measured cost attribution, and Sec.~\ref{sec:threats} records the limitation.

Matched-size Isolated control series probe these warm-series maxima. Between-sample software
cache eviction (the \emph{cold} series) raises the median by at most $+\coldWorstPct\%$
across the fixtures; a co-runner sharing the last-level (L3) cache and stressing one resource
at a time raises the observed maximum by at most $+\coWorstCppPct\%$ for C++, on
\coWorstCppFixture{} with the \coWorstCppMode{} co-runner, and by at most
$+\coWorstRustPct\%$ for Rust, on \coWorstRustFixture{} with the \coWorstRustMode{}
co-runner. Across the warm, cold, and three co-runner series Rust
has the smaller observed maximum in \coRustLowerCells{} of \coPairedCells{} equal-sample
comparisons; these are empirical comparisons, not additive interference bounds.

A unit-cost regression on the counter vector is diagnostic only. The worst regressor
correlation is \regCorr, bootstrap intervals are wide, and the adopted Rust model has a
\regLooMaxRust\% worst leave-one-out error. The implementation comparison therefore rests on the
direct fixture maxima of Table~\ref{tab:tails} rather than fitted coefficients; the models,
coefficients, and intervals are in Appendix~\ref{app:regression}.

\subsection{Per-input work-trace conformance}
\label{sec:eval-tracecert}
\begin{table}[t]
\centering
\caption{Per-input work-trace results. Shape compares pass count, point and neighbor counts, and canonical leaf-ID digests; payload additionally includes leaf parameters. The C++ and Rust traversal counters count different events and are not directly comparable.}
\label{tab:tracecert}
\scriptsize
\setlength{\tabcolsep}{3pt}
\begin{tabular}{lrrrrrr}
\toprule
fixture & passes & shape & payload & score ULP & \flannevals{} & \kdnodes{} \\
\midrule
\emph{search-00} & 31 & exact & different & $12$ & \num{5242829} & \num{12140437} \\
\emph{search-01} & 31 & exact & different & $2$ & \num{5436979} & \num{11070038} \\
\emph{pareto-01} & 31 & exact & different & $4$ & \num{5142501} & \num{12139294} \\
\emph{pareto-02} & 31 & exact & different & $1$ & \num{5030428} & \num{12156351} \\
\emph{dense-nbrs} & 31 & exact & different & $1$ & \num{3632586} & \num{7778334} \\
\emph{max-iters} & 31 & exact & different & $2$ & \num{1082202} & \num{1319894} \\
\emph{cache-hostile} & 4 & exact & different & $0$ & \num{260136} & \num{540048} \\
\emph{subnormal} & 31 & exact & different & $5.3\times 10^{9}$ & \num{563727} & \num{493644} \\
\emph{geom-stress} & 31 & exact & different & $1$ & \num{1895371} & \num{4474819} \\
\emph{shipped-osc} & 31 & exact & different & $4.0\times 10^{7}$ & \num{1435268} & \num{3221237} \\
\emph{real-slowest} & 31 & exact & different & $4.7\times 10^{10}$ & \num{1981459} & \num{3935876} \\
\emph{real-median} & 31 & exact & different & $4.4\times 10^{12}$ & \num{1423933} & \num{1013244} \\
\bottomrule
\end{tabular}
\par\smallskip\footnotesize Line-search entries: 0 on every input (measured); C++ passes = $N_{\mathrm{iter}}{+}1$ exactly on every input. The 6 $P$-sweep instances (not shown) also satisfy the work-trace criterion.
\end{table}

The work-trace conformance check of Sec.~\ref{sec:meth-transfer} was applied to every
frozen fixture and to all \realFrames{} captured real frames. The traced analysis build of
the C++ engine uses an instrumented copy of the engine sources plus a counting FLANN
distance functor; the shipped sources remain byte-identical. It emits a per-pass trace that
the replay compares leg by leg against the Rust engine's mirrored trace. Every quantity is
a deterministic function of the input, so the runs need no controlled timing environment,
and the traced binaries are never used for timing.

Table~\ref{tab:tracecert} shows the fixture leg. The \emph{shape} fields comprise pass
count, per-pass point and neighbor counts, and SHA-256 digests of each point's leaf IDs
after canonical sorting. Shape agrees on all \traceFixtures{} inputs, the line-search loop
is never entered, and the measured C++ pass count equals $N_{\mathrm{iter}}{+}1$ on every
input. This directly establishes equality of the recorded work topology, rather than
inferring it from iterations or output proximity. The independent payload digest, which
also records leaf means and regularized inverse covariances, agrees on
\tracePayloadExactFixtures{} of \traceFixtures{} fixtures. The payload result demonstrates
that shape agreement does not establish equality of all kernel operands.

The C++ neighbor count is the complete result of an uncapped radius search, not a
post-search prefix, whereas the Rust count is the at-most-64 list returned by its capped
search. Shape agreement therefore confirms the bound along the C++ query trajectory.
It does not confirm the counterfactual uncapped count at each Rust query coordinate, which
may differ slightly from its C++ counterpart; that distinction prevents the trace result
from being interpreted as membership in $D$.

The per-pass score agrees to \traceUlpConvMax{} f64 ULP or better on the
well-conditioned fixtures and is bit-identical on \emph{cache-hostile}, but drifts to
$\traceUlpWorstTex$ ULP on the most ill-conditioned geometries. PCL/FLANN returns each
point's neighbors sorted by distance while the port's kd walk returns them in visit order,
so the engines may accumulate the same shape in different orders. This explains why exact
floating-point sums are not required by $E_{\mathrm{shape}}$. SHA-256 makes accidental
digest equality computationally unlikely but does not turn this per-input check into a
formal proof.

The real-data leg applies the same hierarchy to every on-map frame; its exhaustive
partition and timing subsets are reported once in Sec.~\ref{sec:eval-realdata}. Among the
\traceRealWorkGapFrames{} base-only frames, define
$\Delta_{\mathrm{nbr}}=\sumnbr^C-\sumnbr^R$. The median $|\Delta_{\mathrm{nbr}}|$ is
\traceRealWorkGapNbrMed{} and the maximum is \traceRealWorkGapNbrMax, at most
\traceRealWorkGapRelMaxPct\% of a frame's kernel evaluations. Only
\tracePayloadExactOnMap{} on-map frames have an exact payload digest, confirming that
matching shape usually does not imply bit-identical leaf parameters.

\subsection{Heap allocations per align}
\label{sec:eval-alloc}
On fixtures whose radius queries return neighbors, the interposer counts decompose as
${\approx}\allocPerPtPass$ allocations per point per derivative pass. For example,
$\allocWorstPerAlign{} / (31 \cdot 2000) = \allocPerPtPassExact$. This locates the
source in the per-point inner loop: a fresh neighbor
vector constructed inside the parallel loop body, two fresh index/distance vectors per
radius-search call, and approximately eight per-query result containers per point inside PCL/FLANN.
These allocations use the allocator's thread-cache fast path and have low mean cost, but
each is a
potential lock, page fault, or fragmentation stall inside the innermost loop. The Rust
engine hoists all three allocation sites into a pre-reserved workspace and measures zero in
the same harness, consistent with the counting-allocator-verified allocation freedom shown
in Table~\ref{tab:alloc}.

Beyond the hazard reading, these counts provide a separate C++-side consistency check.
The numerator comes from the replay binary containing the unmodified C++ engine via
\texttt{LD\_PRELOAD}, while the denominator
$P \cdot N_{\mathrm{pass}}$ comes from the Rust counters. The ratio is
\allocCertMin--\allocCertMax{} on every fixture whose radius queries return neighbors.
The \allocCertSpreadPct\% spread remains small across geometries whose cost differs by two
orders of magnitude. A single extra C++ derivative pass, which would shift the count by
about 3\% at $N_{\mathrm{pass}} = 31$, would exceed the observed spread. The
one deviation, \emph{subnormal} at \allocCertSparse{}, is itself structural: its
mostly-empty radius searches skip the per-result containers.

\begin{table}[t]
\centering
\caption{C++ allocates ${\approx}\allocPerPtPass$ times per point per derivative pass on neighbor-returning fixtures, whereas the Rust align path allocates zero: heap allocations per align (\texttt{LD\_PRELOAD} interposer).}
\label{tab:alloc}
\footnotesize
\setlength{\tabcolsep}{4pt}
\begin{tabular}{lrrr}
\toprule
fixture & C++ / align & C++ / (pt$\cdot$pass) & Rust / align \\
\midrule
\emph{search-00} & \num{682248} & 11.0 & 0 \\
\emph{search-01} & \num{682198} & 11.0 & 0 \\
\emph{dense-nbrs} & \num{511748} & 11.0 & 0 \\
\emph{max-iters} & \num{409403} & 11.0 & 0 \\
\emph{cache-hostile} & \num{88032} & 11.0 & 0 \\
\emph{subnormal} & \num{292038} & 9.4 & 0 \\
\emph{geom-stress} & \num{511748} & 11.0 & 0 \\
\emph{shipped-osc} & \num{511748} & 11.0 & 0 \\
\bottomrule
\end{tabular}
\par\smallskip\footnotesize The constant ${\approx}11$ per point per pass locates the source in the per-point inner loop (Sec.~\ref{sec:eval-alloc}). Rust: zero, matching the counting-allocator-verified allocation-freedom contract. Allocation counting uses 20 measured aligns per fixture after 5 warmups under Isolated.
\end{table}

\subsection{Scaling with the input size \texorpdfstring{$P$}{P}}
\label{sec:eval-psweep}
We regenerate the Rust-search geometry at six values
$P\in[\psweepPmin{},\psweepPmax{}]$, increasing by a factor of two at each step. The
bounded Rust engine is reconstructed at each point with $P_{\max}=P$; $L_{\max}$ and
$I_{\max}$ remain fixed. Rows above $P=\runtimePMax$ therefore evaluate larger declared
source capacities rather than violating the principal configuration. The counters verify
$N_{\mathrm{iter}}=30$ and $\sumnbr=P\cdot64\cdot31$ at every point, so the common
kernel-evaluation work per source point is invariant. Rust kd traversal per processed point
spans \psweepKdPerPointMin--\psweepKdPerPointMax{} nodes across the six fixtures; it is
measured rather than assumed constant.

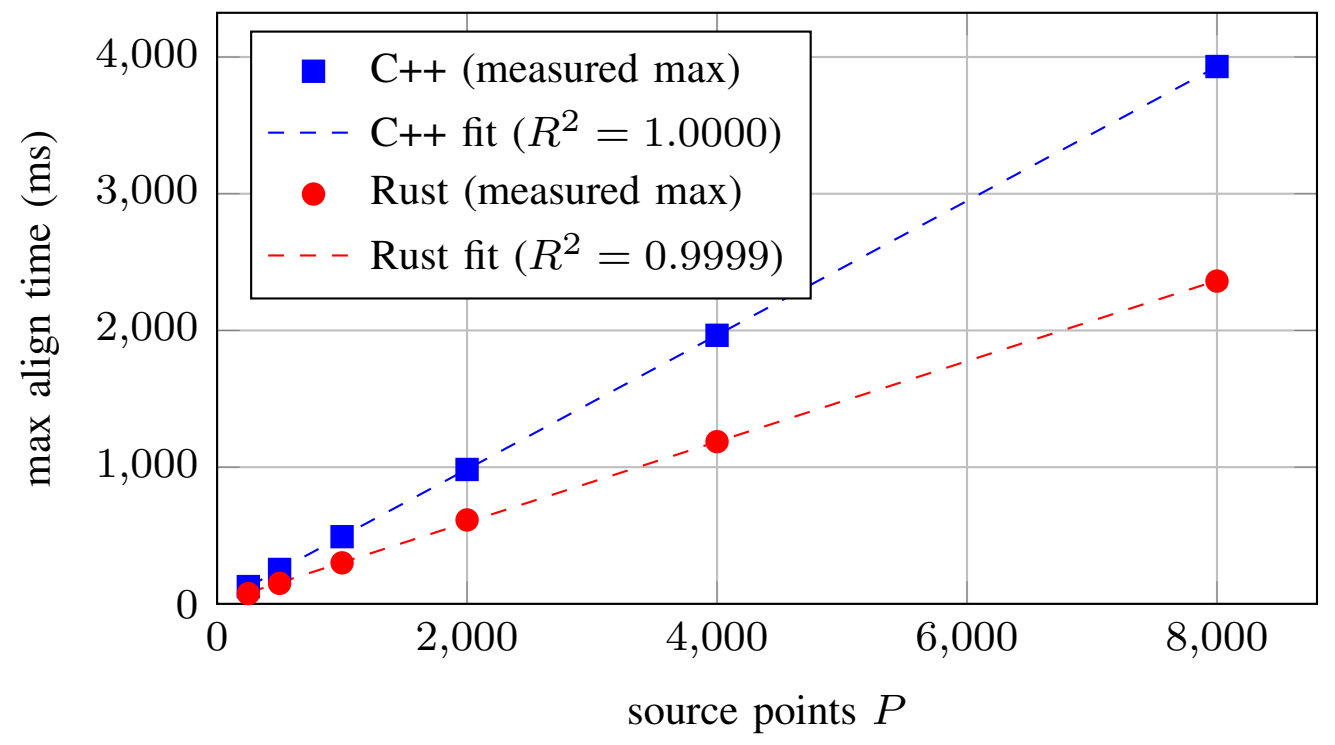
\begin{figure}[t]
\centering
\begin{tikzpicture}
\begin{axis}[
  width=\columnwidth, height=0.62\columnwidth,
  xlabel={source points $P$}, ylabel={max align time (ms)},
  xmin=0, ymin=0, legend pos=north west, legend cell align=left,
  grid=major, tick label style={font=\footnotesize}, label style={font=\footnotesize},
  legend style={font=\footnotesize},
]
\addplot[only marks, mark=square*, color=blue] coordinates { (250,126.34) (500,251.75) (1000,491.65) (2000,982.99) (4000,1964.51) (8000,3930.39) };
\addlegendentry{C++ (measured max)}
\addplot[no marks, color=blue, dashed] coordinates { (250,125.68) (8000,3929.18) };
\addlegendentry{C++ fit ($R^2=1.0000$)}
\addplot[only marks, mark=*, color=red] coordinates { (250,75.07) (500,149.78) (1000,300.09) (2000,613.89) (4000,1186.54) (8000,2361.07) };
\addlegendentry{Rust (measured max)}
\addplot[no marks, color=red, dashed] coordinates { (250,81.34) (8000,2364.69) };
\addlegendentry{Rust fit ($R^2=0.9999$)}
\end{axis}
\end{tikzpicture}
\caption{Observed maximum align latency vs.\ source-point count $P$ on the
Rust-search geometry
(common kernel evaluations per point held invariant; traversal counters measured separately).
Dashed lines are descriptive affine fits
through the six observed maxima, not hard latency bounds.}
\label{fig:psweep}
\end{figure}

Let $T_{\mathrm{obs,max}}(P)$ denote the largest measured alignment time at source-point
count $P$. Linear common-kernel work motivates the descriptive affine model
$T_{\mathrm{obs,max}}(P)=\mathrm{slope}\cdot P+\mathrm{const}$, while traversal,
fixed per-point work, and platform noise remain in the observations. The fit is not a hard
bound.
Table~\ref{tab:psweep} and Fig.~\ref{fig:psweep} show the result: an affine line closely fits
the \emph{max} series, with coefficient of determination $R^2 = \psweepRsqCpp{}$ for C++ and
$\psweepRsqRust{}$ for Rust. The largest absolute residual is
\SI{\psweepResidualCpp}{ms} for C++ at $P=\psweepResidualPCpp{}$ and
\SI{\psweepResidualRust}{ms} for Rust at $P=\psweepResidualPRust{}$; no cache knee is
visible over the tested range. The fitted
observed-maximum slope is
\SI{\psweepSlopeCpp}{\micro\second} for C++ versus
\SI{\psweepSlopeRust}{\micro\second} for Rust
per source point. The slope is a compound empirical cost that includes 64 kernel
evaluations across 31 passes per point, traversal, and fixed per-point work; the experiment
does not identify separate unit costs.

Inverting the fitted observed-maximum line against the 10\,Hz, \SI{100}{ms} budget gives an
\emph{empirical budget crossover} at $P \approx \psweepCrossRust{}$ for Rust and
\psweepCrossCpp{} for C++ under the transferred Rust-search geometry on this host. The Rust
value interpolates within the measured range, whereas the C++ value is a short extrapolation
below its lower endpoint. These point estimates are neither hard admission limits nor WCET
bounds. Table~\ref{tab:psweep-fit} reports the slope and intercept intervals and residuals;
the same protocol is required to obtain a platform-specific crossover on deployment
hardware.

\begin{table}[t]
\centering
\caption{An affine line closely fits the observed maxima of the capacity-parameterized $P$-sweep on the Rust-search geometry (\si{ms}; common kernel work verified: $N_{\mathrm{iter}}{=}30$ and $\sumnbr = P \cdot 64 \cdot 31$ at every $P$).}
\label{tab:psweep}
\footnotesize
\setlength{\tabcolsep}{4pt}
\begin{tabular}{rrrrr}
\toprule
$P$ & \multicolumn{2}{c}{C++ (p50 / max)} & \multicolumn{2}{c}{Rust (p50 / max)} \\
\midrule
\num{250} & 123.8 & 126.3 & 74.7 & 75.1 \\
\num{500} & 246.6 & 251.7 & 148.0 & 149.8 \\
\num{1000} & 487.4 & 491.7 & 295.8 & 300.1 \\
\num{2000} & 972.4 & 983.0 & 587.7 & 613.9 \\
\num{4000} & 1954.6 & 1964.5 & 1172.8 & 1186.5 \\
\num{8000} & 3891.0 & 3930.4 & 2337.7 & 2361.1 \\
\bottomrule
\end{tabular}
\par\smallskip\footnotesize The sweep regenerates the geometry and sets $P_{\max}=P$ at each point. Its $P{=}2000$ row is distinct from \emph{search-00}. The secondary protocol uses three separately booted sessions with 100 measurements per engine, fixture, and session; it does not share Table~\ref{tab:tails}'s 1000-sample cell size.
\end{table}

\subsection{Two-tier input-domain analysis}
\label{sec:eval-twotier}
The adversarial fixtures are constructed stress witnesses for the capped-neighbor engine
API; they are not upper bounds on the bounded-neighbor domain $D$ defined in
Sec.~\ref{sec:meth-transfer}. That domain is the
relevant contract for a \texttt{no\_std} kernel deployment in which inputs cross a trust
boundary without production preprocessing validation. When the production
preprocessing contracts of Sec.~\ref{sec:impl-legal} hold, the reachable input space
shrinks. The input geometry of \emph{geom-stress} satisfies those contracts, but its run
uses non-shipped $\epsilon=10^{-10}$ to force 30 iterations. Its observed maximum is
\SI{\legalWorstMaxRust}{ms} for Rust and \SI{\legalWorstMaxCpp}{ms} for C++. These are
measured factors of \legalWorstRatioRust$\times$ and \legalWorstRatioCpp$\times$ below
the transferred Rust-search fixture in Table~\ref{tab:tails}. These are measured
witness-to-witness gaps under stress configurations, not shipped-configuration performance
ratios.

The production contracts shrink the neighbor factor (Sec.~\ref{sec:implementation}). This
is a \emph{fixture-to-fixture measurement}, not a proven tightening, and it is conditional
on the contracts because the engine does not check them. Table~\ref{tab:preconditions}
lists these assumptions.

The mechanism is the $K$ ceiling: disjoint voxel-aligned map tiles cap leaf multiplicity at
1, capping $K$ at 27 instead of 64. This witness realizes
$\bar{K} = \legalWorstKbar$, and the production random-downsample cap fixes
$P \le 1500$. The cost structure also shifts: kernel evaluations drop
$\twoTierKernelDrop\times$, with $\sumnbr$ falling
from $\searchZeroSumNbrM\times10^6$ to $\legalWorstSumNbrK\times10^3$, but time
drops less, because kd-tree traversal, unaffected by the $K$ ceiling, becomes the dominant
term. The counter reduction illustrates the contraction in modeled work, whereas the timing
gap also reflects fixture geometry and configuration and cannot be attributed solely to
input-contract enforcement. Independently, the shipped-configuration search of
Sec.~\ref{sec:impl-legal} reaches the full 30-iteration cap on \emph{shipped-osc}. It costs
\SI{\legalOscMaxRust}{ms} in Rust and \SI{\legalOscMaxCpp}{ms} in C++, with
$\bar{K} = \legalOscKbar$, establishing cap reachability at the shipped $\epsilon$
(Sec.~\ref{sec:impl-legal}). At the same shipped $\epsilon$, the
\emph{geom-stress} input converges before the iteration cap.

\subsection{Host scaling of contract-geometry stress}
\label{sec:eval-parallel}
\begin{table}[t]
\centering
\caption{Two workers bring the Rust geom-stress run below the \SI{100}{ms} period on this host. Cells are the maximum align time in \si{ms} at non-shipped $\epsilon=10^{-10}$, one pinned worker per isolated core.}
\label{tab:parallel}
\footnotesize
\setlength{\tabcolsep}{4pt}
\begin{tabular}{llrrr}
\toprule
fixture & engine & $k{=}1$ & $k{=}2$ & $k{=}4$ \\
\midrule
\emph{geom-stress} & C++ & 240.8 & 128.1 (1.9$\times$) & 64.2 (3.8$\times$) \\
 & Rust & 122.9 & 61.0 (2.0$\times$) & 31.7 (3.9$\times$) \\
\bottomrule
\end{tabular}
\par\smallskip\footnotesize Rust iteration counts are $k$-invariant; calibration drift: \parCalibDriftPct\% (no throttling).
\end{table}

Table~\ref{tab:parallel} reports host scaling for the \emph{geom-stress} run at
$\epsilon=10^{-10}$. The
serial Rust result costs \SI{\parRustLegalSerialMs}{ms}, missing the
\SI{100}{ms} budget. We therefore ran both engines with $k\in\{1,2,4\}$ workers on $k$ isolated physical
cores, with one pinned worker per core. The Rust engine's Rayon (data-parallelism
library) reduction preserves point order and
is bit-identical to serial for every $k$. At $k=\parRustLegalFitK$, it reduces the witness
to \SI{\parRustLegalFitMaxMs}{ms}. Thus this non-shipped stress run falls below the period
on this host with two cores. This is a throughput observation, not evidence of shipped
deadline compliance or multi-core WCET.

\subsection{Real-data replay: cropped-map and degraded-prior stress}
\label{sec:eval-realdata}
To relate the two tiers to real inputs, we used an environment-gated hook at the engine
entry points to capture each NDT input: the map tile, per-align guess, and source cloud. The
capture contains \realFrames{} aligns collected after production preprocessing during a
57-minute urban drive in İstanbul from the Istanbul Open Dataset~\cite{istanbulopendataset},
whose route deliberately traverses LiDAR-hostile structures (a sub-sea tunnel with large
elevation change, long bridges, dense urban canyons). A byte-exact join of every captured cloud against the
recording shows the capture covers the drive's first $\approx$\realCaptureSpanMin{} minutes, that \realDupClouds{} clouds
were aligned two or three times by concurrent capture consumers (with different guesses),
and that the capture order is not drive order; per-frame statistics therefore weight some
scan geometries two- or three-fold.
Both engines replayed the resulting buffers offline with
cost counters enabled. The primary replay deliberately discards the recorded
production-filter guesses in favor of a frozen track built from NDT odometry with
constant-velocity extrapolation from a global-navigation-satellite-system (GNSS) seed, so
that the guess stream is independent of the production filter. This track represents
degraded-prior operation, including initialization and filter dropouts; a control replay
with the recorded production guesses appears below.

We treat the experiment as a \emph{cropped-map, degraded-prior stress replay}, useful for
exposing map-boundary numerical branching, stressing initialization and dropout priors,
and identifying port-fidelity failures; it is not a representative nominal-localization
workload over the full recording. The offline benchmark map is cropped, and the
reconstructed trajectory exits it after frame \realOnMapSeqMax{}. Only \realOnMap{}
aligns, distributed across \realSegments{} short
segments in this prefix and covering approximately 2\% of the drive, perform meaningful
matching work. We classify a frame as \emph{on-map} when at least one radius search returns
a neighbor, so $\sumnbr>0$. Every search is empty on the remaining \realOffMap{}
\emph{off-map} frames, making those aligns near-no-ops with a median of
\realOffMapIterMed{} iterations and a C++ p50 of approximately
\SI{\realOffMapCppMed}{ms}.

Table~\ref{tab:realdata} separates these populations. Its first panel reports $P$, maximum
$K$, and the partition by base agreement and trace conformance over all frames. The second reports timing over
all \realOnMap{} on-map frames as a system-level observation and over the
\realShapeOnMap{}-frame $E_{\mathrm{shape}}$ subset as the strict trace-conformant
implementation comparison. The all-frame counts provide capture context, whereas the
near-no-op off-map majority does not characterize localization fidelity.

\begin{table}[t]
\centering
\caption{Only the 501 on-map frames perform meaningful matching, where C++ overruns the \SI{100}{ms} budget while Rust never does. Per-frame distributions from the open-loop İstanbul replay under the Replay configuration.}
\label{tab:realdata}
\scriptsize
\setlength{\tabcolsep}{2pt}
\begin{tabular}{llrrrr}
\toprule
population & metric & $n$ & p50 & p99 & max \\
\midrule
\multicolumn{6}{l}{\emph{All-frame contract observations}} \\
all & $P$ (points) & \num{22416} & 1333 & 1500 & 1500 \\
all & max $K$ (per point) & \num{22416} & 0 & 8 & 9 \\
all & $E_{\mathrm{base}}$ & \num{22183} & --- & --- & --- \\
all & $E_{\mathrm{shape}}$ & \num{22178} & --- & --- & --- \\
all & base-only & \num{5} & --- & --- & --- \\
all & outside $E_{\mathrm{base}}$ & \num{233} & --- & --- & --- \\
\midrule \multicolumn{6}{l}{\emph{On-map timing observations}} \\
all on-map & C++ align (\si{ms}) & \num{501} & 106.5 & 129.0 & 133.3 \\
all on-map & Rust align (\si{ms}) & \num{501} & 20.9 & 49.5 & 57.2 \\
$E_{\mathrm{shape}}$ & C++ align (\si{ms}) & \num{263} & 98.5 & 127.6 & 132.2 \\
$E_{\mathrm{shape}}$ & Rust align (\si{ms}) & \num{263} & 19.1 & 44.1 & 56.2 \\
\bottomrule
\end{tabular}
\par\smallskip\footnotesize On-map means at least one radius query returned a neighbor; \num{501} frames qualify. The on-map partition is $E_{\mathrm{shape}}=263$, base-only=5, and outside $E_{\mathrm{base}}$=233. Of the latter, 28 differ in iteration count. Strict cross-language timing uses only $E_{\mathrm{shape}}$; all-on-map timing is a system-level observation.
\end{table}

The input counters are consistent with the preprocessing contracts: $P\le\realPMax=1500$
on all \realFrames{} frames. The per-point neighbor count reaches at most
$\realKmaxMax$, within the geometric bound of 27 (Sec.~\ref{sec:implementation}); real
centroids deviate from voxel centers and bring a ninth leaf into range. Under the
degraded-prior track, the
iteration cap is reached on \realIterCapPct\% of on-map frames. A control replay of the
same capture with its recorded production extended-Kalman-filter (EKF) guesses isolates the
prior's contribution, since the per-frame inputs are identical except for the guess. On the
\realABCommonN{} frames that are
on-map in both replays the cap share drops from \realABCapDegPct\% to \realABCapProdPct\%,
and over the control replay's own \realProdOnMap{} on-map frames it is
\realProdIterCapPct\%; those frames form \realProdOnMapSegments{} boundary-transit segments
spanning the drive's first \realProdOnMapSpanMin{} minutes. Prior quality therefore
shifts the cap share but does not eliminate it: the cap stays routinely reached with
production-quality priors on this cropped single-tile benchmark map, and the closed-loop
tiled-map replay of the same drive in Sec.~\ref{sec:eval-stack} still reaches it on
\stackRustCapMin--\stackRustCapMax\% of aligned frames for Rust and
\stackCppCapMin--\stackCppCapMax\% for C++. Iteration-cap reachability on captured
production-preprocessed geometry thus holds under degraded priors, production priors, and
the closed-loop replay with production-like recovery; the absolute share depends on prior
quality and map configuration.

Under this track, the 10\,Hz deadline is no longer a marginal worst case:
\realOverrunsCpp{} C++ aligns exceed the \SI{100}{ms} budget, whereas Rust has no
overruns. The timing rows report the \emph{Replay} configuration: normal CFS scheduling,
no pinning, one align per frame in production order, and a pinned \SI{3.2}{GHz} reference
clock on all cores. The observed-maximum frame costs \SI{\realCppMax}{ms} in C++ and
\SI{\realRustMax}{ms} in Rust. The overruns represent
\realOverrunsCppOnMapPct\% of the on-map frames, whose C++ median is
\SI{\realCppMedOnMap}{ms}. These Replay observations are not compared numerically with the
primary Isolated maxima because the auxiliary replay uses a separately pinned binary.
The comparison is specific to the recorded reference clock; a production electronic control unit (ECU) may have
different governor and boost behavior. The replay also excludes co-resident Autoware
workloads to isolate align-kernel timing and preserve cross-engine comparability. It
therefore does not establish full-stack deadline compliance. For fixed binaries and inputs, the counter rows do not depend on the timing
configuration.

The conformance levels differ materially. Over all frames,
$|E_{\mathrm{base}}|=\realBaseAll{}$ and
$|E_{\mathrm{shape}}|=\realShapeAll{}$. The exhaustive on-map partition is
\realShapeOnMap{} trace-conformant frames, \realBaseOnly{} base-only frames, and
\realBaseDivergent{} frames outside $E_{\mathrm{base}}$. Thus
\realShapeOnMapPct\% of meaningful matches support the strict work-trace comparison;
\realNotShapeOnMap{} frames, or \realNotShapeOnMapPct\%, do not. The last group contains
\realOutputDivergent{} frames with equal iteration counts but excessive pose or score
differences and \realIterationDivergent{} frames with different iteration counts.
Translation-only synthetic fixtures do not exercise map-tile ordering or rotational
convergence. The evaluated port canonicalizes tile order and preserves the C++ rotation
convention; under that implementation, iteration-count divergence remains on
\realIterationDivergent{} frames, or \realOnMapDivPct\% of on-map frames and
\realDivPctAll\% of all captured frames. Section~\ref{sec:threats} characterizes both
the iteration and equal-iteration output residuals. The traced replay of
Sec.~\ref{sec:eval-tracecert} records their per-engine outputs and work shape.
The \realIterationDivergent{} divergences cluster in \realDivSegments{} of
\realSegments{} on-map segments but do not exhibit unusually large work counts:
$\bar{K}$ spans \realDivKbarMin{}--\realDivKbarMax{}, and the iteration count spans
\realDivIterMin--\realDivIterMax{}.
At the maximum-delta frame Rust stops after 17 iterations while C++ reaches 30, amplifying
a small numerical perturbation to \SI{\traceRealDivergentTransDeltaMax}{m}. Such sensitivity
is not language-specific: C++ library, compiler, contraction, or ISA changes can also perturb
accumulation, eigensolution, or convergence. The open-loop replay did not measure that
C++-to-C++ variability.

The pose deltas in Table~\ref{tab:realdata} measure substitutability rather than localization
error because only one engine runs in production and the replay has no ground truth.

\subsection{Closed-loop stack replay with GNSS recovery}
\label{sec:eval-stack}
Stack-level replay extends the comparison to the full NDT--EKF--GNSS loop with
production-like recovery. We replayed the same 57-minute cloud, vehicle-twist (linear and
angular velocity), and GNSS streams twice per engine (C++--Rust--Rust--C++), replaying
\stackInputFrames{} clouds at \SI{10}{Hz}. Each run starts from one shared frozen initial
pose on a \SI{\stackMapExtent}{m}-extent map whose dynamic tiles are loaded within
\SI{\stackMapRadius}{m} of the pose estimate, under CFS with the deployed
\stackWorkers-worker configuration: \stackWorkers{} OpenMP threads for C++ and a
\stackWorkers-worker Rayon pool for Rust. A watchdog implements the production pose-initializer recovery flow: a GNSS
seed pose, a call to the NDT align service, an EKF reset, and re-enabling of the
localization nodes. It fires after \SI{\stackStarve}{s} without an accepted NDT pose, that
is, one admitted by the EKF's measurement-acceptance gate. We call such a period a
\emph{starvation episode}. Because a recorded drive cannot perform the standstill that
production recovery requires, the watchdog instead dead-reckons the pose returned by the
align service across the multi-second align latency using the live twist stream. As noted
in Sec.~\ref{sec:eval-realdata}, the route traverses LiDAR-hostile structures, so
all rates below are stress-route values.

All \stackReinitTotal{} re-initializations across the four runs succeeded, and no run ever
lost the dynamically loaded map: map tiles remained available on every frame. Starvation
episodes coincide across engines: every run starves at the same \stackSharedEpisodes{}
most difficult route sections (the tunnel approach and one urban section), and
\stackCppOnlyEpisodes{} further sections starve only the C++ runs. Same-engine repeat
bands are narrow and the engine bands do not overlap. \emph{Accepted-output coverage} is
the fraction of input clouds for which the pipeline published an NDT pose that passed the
EKF acceptance gate. It is
\stackCppSurvMin--\stackCppSurvMax\% for C++ versus \stackRustSurvMin--\stackRustSurvMax\%
for Rust. The runs perform \stackCppReinitMin--\stackCppReinitMax{} re-initializations for
C++ versus \stackRustReinitMin--\stackRustReinitMax{} for Rust, with iteration-cap
shares of \stackCppCapMin--\stackCppCapMax\% versus \stackRustCapMin--\stackRustCapMax\%
of aligned frames at a median of \stackIterMedMin--\stackIterMedMax{} iterations. Median
align cost is \stackCppExeMedMin--\stackCppExeMedMax{} versus
\stackRustExeMedMin--\stackRustExeMedMax{} \si{ms}. A \emph{recovery window} is the
interval from a watchdog trigger to the completed EKF reset, during which the align service
contends with the main callback. The only \SI{100}{ms}-budget
violations are \stackCppOverMin--\stackCppOverMax{} C++ frames per run, all inside these
windows. The maxima are
\SI{\stackCppTimeMax}{ms} versus \SI{\stackRustTimeMax}{ms}; no Rust frame exceeds the
budget in any run. These observations are consistent with the lower align cost allowing
the faster engine to tolerate transient match-quality degradations that drive the slower
engine past the starvation threshold. The EKF acceptance-gate decision differs across
languages on \stackQCrossMin--\stackQCrossMax{} of \stackInputFrames{} frames, and the p99
cross-language difference in the EKF innovation (measurement residual) is
\SI{\stackInnovCrossMin}{m}--\SI{\stackInnovCrossMax}{m}; both exceed the same-engine
repeat bands (\stackQRepeatMin--\stackQRepeatMax{} frames;
\SI{\stackInnovRepeatMin}{m}--\SI{\stackInnovRepeatMax}{m}), consistent with a systematic
engine difference rather than replay noise. With two runs per engine and
feedback-induced trajectory divergence, these remain run-level observations.

\subsection{Bridge experiment: Isolated and Replay configurations}
\label{sec:eval-bridge}
A separate bridge experiment measures configuration sensitivity on one byte-identical
benchmark binary. The same five frozen inputs are replayed under both configurations: the transferred
Rust-search fixture, \emph{geom-stress}, \emph{shipped-osc}, and the slowest and
median-cost real frames frozen from the drive capture. They
use the same \SI{3.2}{GHz} reference clock and harness, differing only in the environment.
Table~\ref{tab:bridge} reports medians, maxima, and the descriptive
interference-inflation ratio, defined as the Replay maximum divided by the Isolated maximum.
These ratios apply only to the bridge dataset and are not correction factors for the primary
bounded-build campaign or the auxiliary real-data replay. The synthetic legs use matched sample counts: $n$ denotes the number of measurements per
engine and input, with a pooled three-session $n{=}3000$ campaign in both configurations, so their
max-vs-max ratios are same-$n$; the real-frame legs are $n{=}100$ in both configurations.

\begin{table}[t]
\centering
\caption{The Isolated and Replay configurations are not engine-neutral: C++ maxima are lower under Replay on all five inputs while Rust ratios stay near one. The ratio column is the descriptive Replay-to-Isolated maximum ratio for one byte-identical binary at a fixed \SI{3.2}{GHz} clock.}
\label{tab:bridge}
\footnotesize
\setlength{\tabcolsep}{4pt}
\begin{tabular}{llrrrrr}
\toprule
input & engine & \multicolumn{2}{c}{Isolated (\si{ms})} & \multicolumn{2}{c}{Replay (\si{ms})} & ratio \\ \cmidrule(lr){3-4} \cmidrule(lr){5-6} & & p50 & max & p50 & max & \\
\midrule
\emph{search-00} & cpp & 963.9 & 969.1 & 934.9 & 959.7 & \textbf{0.99} \\
 & rust & 590.9 & 596.4 & 601.9 & 606.1 & \textbf{1.02} \\
\emph{geom-stress} & cpp & 236.3 & 243.4 & 202.6 & 214.0 & \textbf{0.88} \\
 & rust & 120.1 & 120.8 & 120.3 & 120.9 & \textbf{1.00} \\
\emph{shipped-osc} & cpp & 191.1 & 194.2 & 156.9 & 158.7 & \textbf{0.82} \\
 & rust & 64.5 & 64.9 & 64.1 & 64.7 & \textbf{1.00} \\
\emph{real-slowest} & cpp & 166.4 & 166.6 & 132.5 & 132.9 & \textbf{0.80} \\
 & rust & 55.3 & 55.6 & 56.1 & 56.6 & \textbf{1.02} \\
\emph{real-median} & cpp & 141.9 & 143.0 & 109.7 & 109.9 & \textbf{0.77} \\
 & rust & 6.5 & 6.7 & 6.6 & 6.7 & \textbf{1.01} \\
\bottomrule
\end{tabular}
\par\smallskip\footnotesize Iteration counts are identical across configurations on every input. Synthetic legs are matched-$n$ pooled 3-session campaigns ($n{=}3000$) on both configurations (so the max-vs-max ratios are same-$n$); real-frame legs are dedicated controlled sessions ($n{=}100$ each).
\end{table}

The two host configurations are not engine-neutral. For Rust, the Replay-to-Isolated
ratio spans $\bridgeRustInflMin$--$\bridgeRustInflMax\times$ over the five bridge inputs
on the otherwise-idle host, whereas for C++ it spans
$\bridgeCppInflMin$--$\bridgeCppInflMax\times$ and is below one on all five inputs.
Co-resident load is a separate axis probed by the co-runner series of
Sec.~\ref{sec:eval-tails}. The available controls do
not identify a single mechanism, so the ratios are descriptive configuration-transfer
measurements rather than generic scheduler penalties or additive interference bounds.

The cross-engine ordering remains unchanged for the five bridge inputs; the experiment does
not cover the other fixtures in Table~\ref{tab:tails}. On \emph{geom-stress}, the
cross-engine maximum ratio changes from \bridgeGapBLegalWorst{} under Isolated to
\bridgeGapALegalWorst{} under Replay. Thus neither engine's Isolated maximum should be
assumed to transfer by a constant factor outside the evaluated bridge set.

\subsection{On-target counter transfer to AArch64}
\label{sec:eval-raspi4}
\begin{table}[t]
\centering
\caption{Every recorded counter tuple transfers exactly to the AArch64 Bare-metal target, and repeated timing per fixture has sub-percent spread. Device: Raspberry Pi 4 Cortex-A72, \texttt{no\_std} kernel.}
\label{tab:raspi4}
\footnotesize
\setlength{\tabcolsep}{4pt}
\begin{tabular}{lrrrl}
\toprule
fixture & warm p50 & warm max & cold max (\si{ms}) & counters \\
\midrule
\emph{search-00} & 9901.6 & 9907.9 & 9926.0 & exact \\
\emph{search-01} & 6815.8 & 6819.8 & 6823.4 & exact \\
\emph{dense-nbrs} & 7091.6 & 7094.1 & 7094.0 & exact \\
\emph{max-iters} & 461.3 & 462.7 & 462.3 & exact \\
\emph{cache-hostile} & 131.2 & 131.6 & 132.0 & exact \\
\emph{subnormal} & 148.4 & 148.7 & 148.7 & exact \\
\emph{geom-stress} & 1497.8 & 1499.1 & 1500.4 & exact \\
\emph{shipped-osc} & 833.5 & 834.4 & 838.3 & exact \\
\bottomrule
\end{tabular}
\par\smallskip\footnotesize Counters match the host reference for all eight fixtures. A fixed-work calibration spin brackets every series: worst drift \raspiCalibDriftPct\% over the sustained run, with no calibration evidence of throttling. Firmware-default clock; single core active, no interrupts routed to the benchmark task's core.
\end{table}

To validate the cross-ISA leg of the methodology from Sec.~\ref{sec:meth-crossisa} on an
embedded evaluation target, we ran the same frozen fixtures through the engine on a
Raspberry Pi~4 with a Cortex-A72 under the bare-metal \texttt{no\_std} kernel. An async task
parses the embedded fixture bytes, performs one counted align each, and asserts the counters
against the commit-pinned host snapshot. Table~\ref{tab:raspi4} reports the resulting
timing series and counter verdict. Every counter of every fixture transfers
\emph{exactly}, or \raspiExact{} of \raspiTotal{}.

Commit-pinned host expectations, emulator replay, and hardware agree on
\raspiDeviceOk{} of \raspiTotal{} counter tuples. Appendix~\ref{app:crossisa-controls}
describes the controls for fused multiply-add contraction, emulation, and subnormal
arithmetic.

The timing columns of Table~\ref{tab:raspi4} report the \emph{Bare-metal} configuration for
the OS-less target. For each fixture, the series contains
\raspiTimingSamples{} warm and \raspiTimingCold{} cold samples on the warm map and is
bracketed by a fixed-work
calibration spin, the on-target analog of the host campaign's throttle guard and the
relevant hazard over this sustained run. Across approximately 55 minutes, the calibration
drifted at most \raspiCalibDriftPct\%, providing no evidence of clock or thermal throttling.
The worst warm max/p50 difference across all fixtures is
${+}\raspiSpreadWorstPct\%$. The cold series uses an 8\,MiB eviction sweep before every
sample and sits at most ${+}\raspiColdDeltaWorstPct\%$ above the warm maximum, tighter than
the Isolated host series, under no OS, one active core, and a stable observed calibration rate.
The recorded counter tuples transfer exactly, and repeated timing for each fixed tuple has
sub-percent spread. These observations support deriving a
platform-specific unit-cost bound, but do not themselves establish one. In
absolute terms the serial configuration is not schedulable at 10\,Hz on this target:
\emph{search-00} costs \SI{\raspiSearchWarmMaxS}{s}, or $\raspiHostFactor\times$ the
x86 host's Isolated maximum. The non-shipped \emph{geom-stress} run costs
$\raspiLegalWorstBudgetX\times$ the \SI{100}{ms} budget, and Table~\ref{tab:raspi4}
separately reports \emph{shipped-osc}. These observations show that the A72 is inadequate
for the measured serial stress runs; they do not define a production-ECU speed requirement
or a shipped-input worst case. On these fixtures, the A72 maximum is
$\raspiHostFactorMin$--$\raspiHostFactorMax\times$ the x86 maximum; the x86 host itself
also misses the \emph{geom-stress} period by $\legalWorstBudgetXRust\times$, as shown in
Sec.~\ref{sec:eval-twotier}. The protocol of Sec.~\ref{sec:eval-psweep} can be repeated on
a candidate platform to obtain a platform-specific $P$-sweep. Potential mechanisms for
meeting the deadline include reducing $P$, applying
single-instruction multiple-data (SIMD) vectorization, making
algorithmic changes, and parallelizing the kernel. Section~\ref{sec:eval-parallel} measures
approximately
\parRustSpeedupFour$\times$ at four host cores.

\subsection{Tail diagnostic and rejection of extrapolation}
\label{sec:eval-evt}
For each tail fixture we fitted per-session POT/GPD models to $3{\times}1000$ Isolated
samples and applied the full diagnostic gate. The lag-1--10 autocorrelation function (ACF) reaches
\evtAcfWorst{} rather than the ${\sim}0.03$ expected under independence; consequently
standard independent-sample inference is invalid. In addition,
\evtFitFailures{} of \evtFitChecks{} session-level fits fail the in-sample p99.9 check. These
diagnostics invalidate intervals for $q_{10^{-9}}$, the fitted latency quantile at exceedance
probability $10^{-9}$. We
therefore report no extrapolated exceedance probabilities and retain only the empirical
maxima in Table~\ref{tab:tails}. Table~\ref{tab:evt} in Appendix~\ref{app:evt} gives the
fixture-level diagnostics solely to document this rejection; plausible-looking fits do not
override failed assumptions.

\section{Threats to Validity}
\label{sec:threats}

\subsection{Measurement and coverage}
Isolated timing uses a pinned core, \envSessions{} randomized sessions from distinct boots,
separate warm, cold, and co-runner series, and a calibration guard that excludes cells with
more than 10\% sustained speed drift. Across fixture--engine cells, the largest
cross-session median spread is \sessMedSpreadWorstPct\%. The largest within-fixture range
of session-level Rust/C++ median ratios is \sessRatioSpreadWorst{}. Nevertheless, all timing
comes from one machine. The campaign compares complete deployable implementations under recorded
release settings, not languages in isolation; compiler flags have no exact cross-toolchain
mapping, and allocators, layouts, kd-trees, libraries, and compiler choices all contribute.
The open-loop real-data timing replay measures each captured frame once, the closed-loop
replay carries only two runs per engine, and parallel results omit worker scheduling and
interference, so none supports a multi-core WCET claim. The unified campaign and
$P$-sweep are the primary bounded-build timing evidence. Bridge, parallel, subnormal-control,
real-data (open-loop and closed-loop), and AArch64 results are artifact-pinned auxiliary
experiments; they are not pooled
with the primary campaign, and their measurements do not characterize its binary.

Search covers a parameterized generator, not all inputs. Kernel-evaluation saturation is
exact, but the traversal objective is the Rust-specific kd-node count and has only its
$O(N_{\mathrm{leaves}})$ structural cap; a slower Rust geometry may exist. The counter vector also omits empty-query frequency, fixed per-point overhead,
solver branches, and numerical regime. No independent search uses the C++ FLANN distance or
plane-evaluation counters, so C++ timing evidence is limited to observations over the
transferred fixture set. The real-data evidence covers one
vehicle, route, and cropped map: the primary degraded-track replay performs meaningful
matching on only \realOnMap{} on-map frames in \realSegments{} short segments, and the
production-guess control replay on \realProdOnMap{} boundary-transit frames. The remaining
frames are near-no-op aligns and therefore do not characterize localization
work over the full recording. The degraded-prior track uses offline NDT odometry rather than
the production filter, so it does not represent nominal cruising or closed-loop behavior;
the control replay restores the recorded production guesses but remains open-loop on the
same cropped map.
The NDT--EKF--GNSS replay includes feedback and GNSS re-initialization but lacks ground
truth, vehicle dynamics, and co-resident workloads. Its watchdog replaces production's
standstill requirement with twist-based motion compensation because a recorded drive
cannot stop for recovery. The \SI{\stackStarve}{s} starvation threshold was selected using
the first C++ run and held fixed for all four runs; the results may be sensitive to this
choice. The recovery statistics of Sec.~\ref{sec:eval-stack} are stress-route values and
carry two runs per engine. Same-engine repetitions also diverge, preventing whole-drive
language attribution; environment-gated EKF tracing may itself perturb scheduling.

\subsection{Cross-language and cross-ISA validity}
For the evaluated port, iteration-count divergence in the primary degraded-track replay
occurs on \realIterationDivergent{} of \realFrames{} frames, all among the \realOnMap{}
on-map set, or \realOnMapDivPct\%; the production-guess control replay shows a comparable
on-map rate of \realProdIterationDivergent{} of \realProdOnMap{} frames, likewise all
on-map. The stricter $E_{\mathrm{base}}$ criterion excludes
\realBaseDivergent{} on-map frames: \realOutputDivergent{} have equal iterations but exceed
at least one output threshold, and \realIterationDivergent{} take different iteration
paths. Residual differences are consistent with glibc versus \texttt{libm}, eigensolver
behavior, and neighbor accumulation order.

Work-trace claims apply only to $E_{\mathrm{shape}}$; \realBaseOnly{} base-only frames
still differ by up to \traceRealWorkGapRelMaxPct\% in kernel evaluations. The canonical
SHA-256 shape digest records leaf identity but not leaf parameters. The separate payload
digest includes means and regularized inverse covariances, yet agrees on only
\tracePayloadExactFixtures{} synthetic fixtures and \tracePayloadExactOnMap{} on-map
frames. Consequently, neither digest establishes full semantic equivalence. SHA-256
collisions remain computationally possible. The C++ trace records complete uncapped
neighbor lists, so shape agreement confirms at most \maxnn{} leaves at the C++ query
coordinates. The Rust search detects a 65th leaf at its own coordinate; a
\runtimeValidationFrames{}-frame validation replay triggered no such diagnostic. This
does not remove the coordinate difference or retroactively add the diagnostic to the
cross-language trace relation.

The Rust stream targets ISA stability through software \texttt{libm} and no fused
multiply-add (FMA) contraction; all \raspiTotal{} frozen fixtures match on the AArch64
target. This is a per-fixture check, not a general ISA proof. Expected counters and fixture
bytes must remain commit-pinned. The C++ leg remains x86/Linux-specific because it depends
on PCL and ROS.

\subsection{Safety and contract scope}
Rust reduces the verification surface but is not required for bounded execution; an
appropriately redesigned C++ engine could impose the same loop, allocation, recursion, and
neighbor bounds. Safe Rust discharges memory-safety and data-race-freedom obligations, but
the C ABI, unsafe internals, and trusted dependencies remain in scope. Likewise,
\texttt{no\_std} removes a library dependency rather than guaranteeing real-time behavior.

\begin{table}[t]
\centering
\caption{Preconditions of the conditional Rust work envelope~\eqref{eq:rust-envelope}: each limit, its enforcer, and the
behavior on violation. ``assumed'' rows are not runtime-checked by the engine.}
\label{tab:preconditions}
\scriptsize
\setlength{\tabcolsep}{2pt}
\begin{tabular}{@{}p{0.18\columnwidth}p{0.12\columnwidth}p{0.24\columnwidth}p{0.36\columnwidth}@{}}
\toprule
quantity & limit & enforced by & on violation \\
\midrule
$P$ & \runtimePMax & engine limit & return source-limit error \\
$N_{\mathrm{iter}}$ & 30 & engine and loop guards & reject larger setting \\
line-search passes & 0 extra & shipped flag off & outside proposition if enabled \\
$K$ (Rust) & 64 & hard search cap & warn; force non-converged \\
$K$ (C++) & none & --- & grows with tile overlap \\
$K$ (production input) & 27 & preprocessing contract (assumed) & production-input bound invalid \\
$N_{\mathrm{leaves}}$ & \runtimeLeavesMax & map finalization & reject; retain active map \\
kd query stack & 64 entries & build-depth guard & return map-build error \\
serial scratch & \runtimeScratchBytes{} B & startup reservation & reject undersized scratch \\
integer products & counter range & checked construction & return arithmetic error \\
numeric inputs & finite & align entry checks & return non-finite error \\
\bottomrule
\end{tabular}
\end{table}

Table~\ref{tab:preconditions} distinguishes enforced limits from assumptions.
Engine-configuration stress fixtures are API-legal but need not satisfy production
preprocessing contracts. Those contracts include tile geometry, downsampling, and the
tighter operational $P\le1500$ cap. The evaluated Rust engine
enforces $P\le\runtimePMax$ and $N_{\mathrm{leaves}}\le\runtimeLeavesMax$, but it does
not verify the tighter preprocessing or geometry conditions. Production must
validate static map conditions at load time and per-scan conditions before alignment, with a
defined rejection or fallback. Validator execution must be bounded and included in the
10\,Hz deadline analysis.

No searched contract-conforming input combines $K=8$ with 30 shipped-parameter iterations:
\emph{geom-stress} pins the latter through non-shipped $\epsilon$, while
\emph{shipped-osc} establishes a cap-reaching witness at shipped settings. The analytic $K\le27$
ceiling has no witness and assumes boundary and floating-point behavior. A
\legalKSearchSeeds{}-seed contract-conforming search reaches only
$K=\legalKSearchMax$; this is evidence, not proof, that reachable $K$ stays below the
ceiling. The affine $P$ fit is validated only for the capacity-parameterized family
$P\in[\psweepPmin{},\psweepPmax{}]$ on the host; each point uses $P_{\max}=P$, so the
rows above \runtimePMax{} do not belong to the principal source-capacity configuration.
Allocation freedom holds only inside the capacity reserved for the selected configuration and
removes one kernel-level latency path. It does not bound interrupts, system-management
interrupts, page faults, scheduler delays,
or memory-hierarchy interference on the Linux host.

\subsection{Evidence strength}
\label{sec:threats-epistemic}
Loop guards, the neighbor cap, reserved buffers, and the fixed-depth stack are enforced by
construction. Linting excludes explicit panic paths in project-owned code, while property,
allocation, differential, and replay tests provide dynamic evidence. We claim no
machine-checked formal proof. The Rust envelope's maximum for the common
kernel-evaluation term has an exact saturation witness, whereas neither a Rust traversal
maximum nor a C++ traversal extremum is established.

All times are platform-specific observations. The tail diagnostics find serial dependence
reaching $|\mathrm{ACF}|=\evtAcfWorst$; moreover, \evtFitFailures{} of
\evtFitChecks{} fits fail the in-sample adequacy check. Extrapolation is therefore
withheld. Separately, the
bridge experiment associates
the controlled configuration with a \SI{\bridgeCppTaxMinMs}--\SI{\bridgeCppTaxMaxMs}{ms}
unresolved C++ shift. The result is therefore a structurally enforced and property-tested
work envelope plus observed maxima, not a certified hard timing bound. On a predictable
target, a defensible per-operation bound could combine with the work envelope to establish
an analytical WCET.

\section{Related Work}
\label{sec:related}

Static WCET analyzers derive certified time bounds from program-flow and hardware
models~\cite{ferdinand2001ait,wilhelm2008wcet}. Parametric WCET analysis retains
input-dependent quantities in symbolic formulas~\cite{bygde2008parametric,huber2012parametric};
static analysis is therefore not limited to multiplying independent maxima. In this NDT
kernel, however, Newton iterations, neighbor multiplicity, and kd-tree traversal are jointly
induced by floating-point geometry, while a useful bound must be tied to a realizable input.
Our parameterized work envelope is not a processor-time analysis: it enforces and
property-tests structural limits, constructs witnesses for selected terms, and measures them.
It complements rather than replaces certified static hardware analysis.

Measurement-based probabilistic timing analysis (MBPTA) applies extreme value theory (EVT)
to execution-time samples to infer probabilistic worst-case execution time
(\pwcet{})~\cite{edgar2001gumbel,cucu2012mbpta,cazorla2019mbpta}. Its validity depends on both
test coverage~\cite{law2016coverage} and the statistical structure of the measurements.
Lima et al.~\cite{lima2016careful} discuss failure modes in applying EVT to execution times;
Guet et al.~\cite{guet2017representativity} analyze serial dependence and multi-mode tasks,
while Bozhko et al.~\cite{bozhko2023pwcet} formalize the guarantees ascribed to \pwcet{}.
Classical dependent-EVT work formalizes local dependence and the extremal
index~\cite{leadbetter1983extremes}; threshold-exceedance and declustering methods address
serial dependence and clustered extremes~\cite{davison1990threshold,ferro2003clusters}.
These concerns apply directly here because geometry induces distinct modes and the frozen
series exhibit substantial autocorrelation. We address input coverage by constructing stress
fixtures and ranking them with deterministic counters, but the generator still covers only a
parameterized family. The POT/GPD fits fail the predeclared dependence or adequacy gates in
Sec.~\ref{sec:eval-evt}; consequently, we report observed maxima rather than an extrapolated
\pwcet{} estimate.

Search-based timing-stress generation dates to Wegener et
al.~\cite{wegener1997testing,wegener2001comparison}, who use measured execution time and
describe evolutionary testing as complementary to static analysis. Because wall-clock
fitness is noisy and platform-specific, we rank fixtures with a language-common kernel-work
counter and a Rust-specific traversal counter. For a fixed Rust implementation, this ranking
is timing-independent. Only the common counter is shown to reach its structural ceiling.
Replaying Rust-search fixtures on C++ tests per-input work-trace conformance, but establishes
neither a C++ traversal extremum nor a time bound.

ROS~2 chain analyses use callback WCETs to bound end-to-end response
times~\cite{casini2019ros2}. Multi-executor analysis covers reaction time and data age in an
autonomous-driving case study~\cite{teper2024ros2}, while a recent survey spans executor,
communication, tracing, and accelerator timing~\cite{casini2026ros2survey}. ROS~2
autonomous-driving architectures and ROS~2--Adaptive AUTOSAR integration add vehicle-level
deployment context~\cite{bonci2026ros2av,hong2024autosar}. We work below that level, excluding
map loading, scheduling, communication, and downstream callbacks. Our callback evidence is
conditional and platform-specific, not a certified callback time or 10-Hz end-to-end
guarantee.

Rust's ownership model supports resource-constrained systems such as Tock
OS~\cite{levy2017tock}, while RustBelt~\cite{jung2018rustbelt} gives machine-checked
foundations for core language safety claims. Safe Rust narrows memory-safety and data-race
obligations in our safe core; it does not establish WCET. Panic-, allocation-, recursion-,
and loop-boundedness remain separate, and unsafe code, the C ABI, and dependencies remain
trusted. Here the port serves both as an engine and as a controlled instrumentation surface.

NDT registration originates with Biber and Stra{\ss}er~\cite{biber2003ndt} and was extended
to 3D by Magnusson~\cite{magnusson2009thesis}. Autoware's production matcher~\cite{autoware}
is a multigrid derivative of the parallel \texttt{ndt\_omp} implementation~\cite{ndtomp}.
Prior work demonstrates real-time NDT pose correction and reports timing across map
resolutions~\cite{lim2020realtimendt}; it primarily evaluates throughput, accuracy, and
robustness. LIO-SAM and FAST-LIO2 likewise emphasize accuracy, rate, and computational
efficiency in LiDAR--inertial localization~\cite{shan2020liosam,xu2022fastlio2}. We instead
study conditional structural limits, stress witnesses, and cross-language work traces for one industrial align kernel, without proposing a localization
method or a hard deadline guarantee. We are not aware of a published WCET-oriented NDT
analysis with this combination.

\section{Conclusion}
\label{sec:conclusion}

The case study separates analysis of a data-dependent industrial kernel into three
obligations: enforce and test a work envelope, generate stress witnesses with deterministic
counters, and measure frozen inputs. A domain-informed construction attains the common
kernel-count ceiling, while two archived candidates exercise high Rust traversal counts at
lower common-kernel work. The port removes the per-point allocation hazard within
reserved capacities. Rust has the
smaller observed maximum on every warm-series fixture and in
\coRustLowerCells{} of \coPairedCells{} equal-sample comparisons across the warm, cold,
and co-runner series. These timing results remain empirical.

Per-input traces compare a precisely defined work shape without equating leaf payloads,
implementation-specific traversal, or latency. All frozen synthetic fixtures satisfy the
work-trace criterion, although their payload digests differ. The auxiliary cropped-map,
degraded-prior replay provides a more restrictive result: only \realOnMap{} frames perform
meaningful matching, \realShapeOnMap{} satisfy both the base and work-trace criteria, and
\realBaseDivergent{} fall outside the base criterion. Among the latter,
\realIterationDivergent{} take different iteration paths. The replay therefore exposes
port-fidelity and map-boundary sensitivity rather than nominal operation.

Rust is not necessary for these bounds. Its safe core narrows memory-safety and race
obligations so validation can focus on equivalence, numerics, work, and timing; FFI, unsafe
code, and dependencies remain trusted.

The auxiliary NDT--EKF--GNSS replay extends the evidence to feedback and recovery on a
stress route: both engines starve at the same most difficult sections, and every GNSS
re-initialization succeeds. The engine bands stay disjoint, with accepted-output coverage
\stackCppSurvMin--\stackCppSurvMax\% for C++ versus
\stackRustSurvMin--\stackRustSurvMax\% for Rust, and the only \SI{100}{ms} overruns
are C++ frames inside recovery windows. It does not establish production readiness.
Operational evaluation requires the production ECU, ground truth, co-resident loads, and
end-to-end \SI{10}{Hz} acceptance-gate and innovation measurements. Host parallelism places the
non-shipped \emph{geom-stress} run below the period with two cores; it establishes neither
shipped deadline compliance nor multi-core WCET. Stronger timing claims require either a
deployment cost bound or an independence-preserving measurement protocol.

\section*{Acknowledgments}

This paper is based on results obtained from a project subsidized by the New Energy and Industrial Technology Development Organization (NEDO).

\appendices
\section{NDT Align: Pseudocode and Cost-Model Correspondence}
\label{app:pseudocode}
Algorithm~\ref{alg:ndt-align} states the shipped \ndt{} align kernel at the specification
level, in Python-like notation, so that each
factor of the execution-time decomposition~\eqref{eq:decomp} maps to a named step. It is
notation, not the shipped C++/Rust implementation.
Line-search derivative refinement is disabled in the shipped configuration, so the kernel
performs one initial \texttt{deriv\_pass} plus one per Newton step, giving
$N_{\mathrm{pass}}=N_{\mathrm{iter}}+1$.

\begin{algorithm}[t]
\caption{The shipped serial \ndt{} align kernel (Python-style pseudocode). \texttt{cfg}
holds $I_{\max}$, $\epsilon$ (\texttt{trans\_epsilon}), and the step-size cap.}
\label{alg:ndt-align}
\begin{lstlisting}[style=pypseudo]
def deriv_pass(map, source, x):
    score, g, H = 0, 0, 0
    for p in source:                # $P$ points
        q = transform(x, p)
        # radius_search returns $K(p)\le\maxnn$ leaves
        neighbors = radius_search(map.tree, q)
        for leaf in neighbors:      # per point-leaf term
            score, g, H += leaf_term(q, leaf)
    return score, g, H

def ndt_align(map, source, guess):
    x = euler_of(guess); i = 0
    # initial derivative pass
    score, g, H = deriv_pass(map, source, x)
    while (delta := solve_newton(H, -g)) is not None:
        # step length, then pose update
        a = clamp(norm(delta), $\epsilon$/2, cfg.step)
        x = x + a * unit(delta)
        # one derivative pass per Newton step
        score, g, H = deriv_pass(map, source, x)
        i += 1
        if i >= $I_{\max}$ or a < $\epsilon$: break
    return x, i                     # $i = N_{\mathrm{iter}}$
\end{lstlisting}
\end{algorithm}

Table~\ref{tab:eqmap} maps every symbol of~\eqref{eq:decomp} to its step. The bounds
of Sec.~\ref{sec:decomposition} follow directly: \texttt{deriv\_pass} runs
$N_{\mathrm{pass}}=N_{\mathrm{iter}}+1$ times, its point loop $P$ times, and its leaf loop
$K(p)\le\maxnn$ times, while \texttt{solve\_newton} and the pose update run once per
Newton iteration ($N_{\mathrm{iter}}\le I_{\max}$ times). The pseudocode is
implementation-agnostic; the C++ and Rust engines realize the same work shape (the
per-pass structure verified in Sec.~\ref{sec:eval-tracecert}).

\begin{table}[t]
\centering
\caption{Correspondence between the decomposition~\eqref{eq:decomp} and
Algorithm~\ref{alg:ndt-align}.}
\label{tab:eqmap}
\scriptsize
\setlength{\tabcolsep}{3pt}
\begin{tabular}{@{}lll@{}}
\toprule
Term & Step in Alg.~\ref{alg:ndt-align} & Meaning \\
\midrule
$N_{\mathrm{pass}}$ & \texttt{deriv\_pass} calls & derivative-pass count ($=N_{\mathrm{iter}}{+}1$) \\
$P$ & \texttt{for p in source} & source points per pass \\
$T_{\mathrm{search}}(p)$ & \texttt{radius\_search} & one radius search \\
$K(p)$ & \texttt{len(neighbors)} & leaves returned ($\le\maxnn$) \\
$T_{\mathrm{kernel}}$ & \texttt{for leaf in neighbors} & point--leaf score/grad/Hessian \\
$N_{\mathrm{iter}}$ & \texttt{while} loop & Newton updates ($\le I_{\max}$) \\
$T_{\mathrm{solve}}$ & \texttt{solve\_newton} + update & Newton solve + pose update \\
\bottomrule
\end{tabular}
\end{table}

\section{Supplementary Diagnostics}

\subsection{Regression details}
\label{app:regression}
For each engine $e\in\{C,R\}$ and evaluated input, let $Y_e$ be the empirical p50
alignment time, $M$ the number of processed point--pass pairs, $S=\sumnbr$ the kernel-evaluation
count, and $Q_C=\flannevals$ and $Q_R=\kdnodes$ the engine-specific traversal counts. The
two-term model is $Y_e=a_eS+b_eQ_e+c_e$; the four-term model adds
$c_{\mathrm{pt},e}M$. Thus $a_e$, $b_e$, $c_e$, and $c_{\mathrm{pt},e}$ are fitted
kernel, traversal, intercept, and per-processed-point coefficients, respectively.
Table~\ref{tab:regression} reports both models over $n=\regN$ input observations per
engine. Regressor correlation reaches \regCorr, and wide intervals plus leave-one-out error
make coefficient-level attribution unsupported.
\begin{table}[t]
\centering
\caption{Wide intervals and regressor correlation up to 0.99 leave coefficient-level cost attribution unsupported. Brackets are 95\% bootstrap CIs; a star marks the adopted diagnostic model per engine.}
\label{tab:regression}
\scriptsize
\setlength{\tabcolsep}{1.2pt}
\begin{tabular}{llrrrrr}
\toprule
engine & model & $c_{\mathrm{pt},e}$ (\si{\micro\second}/pt) & $a_e$ (\si{ns}/kern.) & $b_e$ (\si{ns}/event) & $c_e$ (\si{ms}) & $R^2$ \\
\midrule
C++ & 2-term & --- & 141.0 [57,\,172] & 77.1 [52,\,143] & 12.1 [-13,\,50] & 0.9989 \\
 & 4-term\,$\star$ & 2.95 [-1.1,\,4.6] & 164.9 [47,\,218] & 26.2 [-31,\,161] & -10.2 [-28,\,7] & 0.9997 \\
Rust & 2-term\,$\star$ & --- & 55.6 [17,\,95] & 30.4 [17,\,43] & -3.4 [-22,\,5] & 0.9988 \\
 & 4-term & -1.37 [-3.2,\,-0.2] & 40.8 [3,\,85] & 41.7 [22,\,63] & 9.0 [-3,\,19] & 0.9993 \\
\bottomrule
\end{tabular}
\par\smallskip\footnotesize Ten fixture points use the unified three-boot timing campaign; the six $P$-sweep points use a separate three-boot, 100-sample-per-session Isolated sweep.
\end{table}

\subsection{\texorpdfstring{$P$}{P}-sweep fit diagnostics}
\label{app:psweep-fit}
Table~\ref{tab:psweep-fit} reports descriptive uncertainty for the affine line through the
six observed maxima. The intervals rely on ordinary least-squares (OLS) residual assumptions and
do not bound unobserved maxima or worst-case latency.
\begin{table}[t]
\centering
\caption{Descriptive affine fit to the six observed maxima; brackets are 95\% OLS line-fit intervals, not confidence bounds on worst-case latency.}
\label{tab:psweep-fit}
\scriptsize
\setlength{\tabcolsep}{2pt}
\begin{tabular}{lrrrr}
\toprule
engine & slope (\si{\micro\second}/pt) & intercept (\si{ms}) & worst residual (\si{ms}) & $R^2$ \\
\midrule
C++ & 490.8 [489.8,\,491.8] & 3.0 [-0.8,\,6.7] & +3.4 ($P=500$) & 1.0000 \\
Rust & 294.6 [290.6,\,298.7] & 7.7 [-7.5,\,22.9] & +17.0 ($P=2000$) & 0.9999 \\
\bottomrule
\end{tabular}
\end{table}

\subsection{Rejected tail-model diagnostics}
\label{app:evt}
Table~\ref{tab:evt} reports fitted-tail diagnostics only to document rejection: $\xi$
denotes the GPD shape parameter. Serial dependence invalidates inferential intervals, and the
in-sample column records fits that fail even before extrapolation. A pooled Gumbel estimate that lies only
\evtGumbelExtraPct\% above the observed maximum illustrates why fit appearance is insufficient.
\begin{table}[t]
\centering
\caption{Rejected per-session POT/GPD tail models ($n{=}1000$/session; 95\% threshold). Dependence or in-sample fit failure invalidates extrapolation; values are diagnostic outputs, not timing estimates. Times in \si{ms}.}
\label{tab:evt}
\footnotesize
\setlength{\tabcolsep}{3pt}
\begin{tabular}{llrrrr}
\toprule
fixture & engine & max & $|\mathrm{ACF}|_{\max}$ & $\xi$ range & fit failures \\
\midrule
\emph{search-00} & C++ & 988.2 & 0.12--0.38 & -0.26..0.54 & 0/3 \\
 & Rust & 588.1 & 0.16--0.42 & -0.43..0.12 & 0/3 \\
\emph{geom-stress} & C++ & 239.8 & 0.48--0.58 & 0.82..1.28 & 1/3 \\
 & Rust & 122.2 & 0.50--0.53 & 0.44..0.68 & 0/3 \\
\emph{shipped-osc} & C++ & 193.0 & 0.63--0.74 & -0.18..0.22 & 0/3 \\
 & Rust & 65.8 & 0.30--0.37 & -0.07..0.51 & 0/3 \\
\bottomrule
\end{tabular}
\par\smallskip\footnotesize ACF: largest lag-1--10 magnitude over three sessions (i.i.d.: ${\sim}0.03$). Fit failures count sessions whose fitted p99.9 differs from the empirical p99.9 by more than three observed tail widths. \emph{geom-stress} uses production-contract geometry with non-shipped $\epsilon$; \emph{shipped-osc} also fixes shipped $\epsilon$.
\end{table}

\subsection{Cross-ISA consistency controls}
\label{app:crossisa-controls}
The cross-ISA check uses commit-pinned fixture bytes and host counter tuples. Rust
software \texttt{libm} fixes the elementary-function implementation, fused multiply-add
contraction is disabled, and both QEMU and AArch64 hardware replay the same serialized
inputs. Agreement is checked on the complete counter tuple rather than inferred from host
results.

\bibliographystyle{IEEEtran}
\bibliography{references}

\end{document}